\documentclass[12pt]{iopart}
\usepackage[utf8]{inputenc}
\usepackage[pdftex]{graphicx}
\usepackage{iopams}
\usepackage[amssymb]{SIunits}
\usepackage{cleveref}
\usepackage{etoolbox}
\crefname{equation}{}{}
\patchcmd{\numparts}{\addtocounter{equation}{1}}{\refstepcounter{equation}}{}{}
\newcommand{\abs}[1]{\ensuremath{\left\vert #1 \right\vert}}
\newcommand{\erw}[1]{\ensuremath{\langle #1 \rangle}}
\newcommand{\order}[1]{\Or(#1)}
\newcommand{\bleq}{\ensuremath{\mathrel{\phantom{=}}}}
\newcommand{\nnl}{\nonumber\\}
\newcommand{\bra}[1]{\langle #1 \hspace{-2pt} \mid}
\newcommand{\ket}[1]{\mid \hspace{-1pt} #1 \rangle}
\renewcommand{\vec}[1]{\mathrm{\mathbf{#1}}}
\newcommand{\up}{\ket{\,\uparrow}}
\newcommand{\down}{\ket{\,\downarrow}}
\newcommand{\bup}{\bra{\uparrow\,}}
\newcommand{\bdown}{\bra{\downarrow\,}}
\newcommand{\ud}{{\uparrow\downarrow}}
\newcommand{\wsn}{\omega_\mathrm{SN}}
\newcommand{\Wsn}{\Omega_\mathrm{SN}}
\newcommand{\taua}{\tau_\mathrm{acc}}

\begin{document}
\title{Dephasing and inhibition of spin interference from semi-classical self-gravitation}
\author{Andr\'e Gro{\ss}ardt}
\address{Institute for Theoretical Physics, Friedrich Schiller University Jena, Fr\"obelstieg 1, 07743 Jena, Germany}
\ead{andre.grossardt@uni-jena.de}

\begin{abstract}
We present a detailed derivation of a model to study effects of self-gravitation from semi-classical gravity, described by the Schrödinger-Newton equation, employing spin superposition states in inhomogeneous magnetic fields, as proposed recently for experiments searching for gravity induced entanglement. Approximations for the experimentally relevant limits are discussed. Results suggest that spin interferometry could provide a more accessible route towards an experimental test of quantum aspects of gravity than both previous proposals to test semi-classical gravity and the observation of gravitational spin entanglement.

\noindent{\it Keywords\/}: semi-classical gravity, Schrödinger-Newton equation, experimental tests of quantum gravity
\end{abstract}

\section{Introduction}
Although there is no consensus about the correct quantum theory of gravity at high energies, there is a prevalent believe that the gravitational field should be quantized in \emph{some} way. For low energies, one then assumes perturbative quantum gravity to apply as an effective field theory, and in nonrelativistic quantum mechanics the Newtonian potential between particles can be treated in complete analogy to the electromagnetic Coulomb potential.

Nonetheless, there is no empirical data how quantum matter acts as a source of gravity, and arguments that quantization would be necessary for theoretical consistency~\cite{eppleyNecessityQuantizingGravitational1977,pageIndirectEvidenceQuantum1981} are inconclusive~\cite{mattinglyQuantumGravityNecessary2005,mattinglyWhyEppleyHannah2006,albersMeasurementAnalysisQuantum2008}. A fundamentally semi-classical approach in which spacetime retains its general relativistic, geometric properties~\cite{rosenfeldQuantizationFields1963,kibbleSemiClassicalTheoryGravity1981,penroseGravitizationQuantumMechanics2014,tilloySourcingSemiclassicalGravity2016}, therefore, remains, if not plausible, at least possible.

Driven by the tremendous progress of quantum experiments with mesoscopic systems, the feasibility of experimental tests of this possibility has been explored with growing attention, a primary route being direct tests of the Schrödinger-Newton (SN) equation~\cite{carlipQuantumGravityNecessary2008,giuliniGravitationallyInducedInhibitions2011,yangMacroscopicQuantumMechanics2013,grossardtOptomechanicalTestSchrodingerNewton2016} which follows as the nonrelativistic limit from the semi-classical Einstein equations as a theory in which curvature of a classical spacetime is sourced by the expectation value of the stress-energy operator of the quantum matter fields~\cite{giuliniSchrodingerNewtonEquationNonrelativistic2012,bahramiSchrodingerNewtonEquationIts2014}.
The SN equation comprises a nonlinear, self-gravitational potential,
\begin{equation}\label{eqn:pot-sn-self}
V_\mathrm{self} = -G m^2 \int \rmd^3 r' \frac{\abs{\psi(t,\vec r')}^2}{\abs{\vec r - \vec r'}}
\end{equation}
in the case of a single point particle, which predicts an inhibition of the dispersion of a free wave packet~\cite{carlipQuantumGravityNecessary2008,giuliniGravitationallyInducedInhibitions2011} as well as modifications for both the dynamics~\cite{yangMacroscopicQuantumMechanics2013} and spectrum~\cite{grossardtOptomechanicalTestSchrodingerNewton2016} of mesoscopic particles in a harmonic trapping potential.

More recently, it has been proposed to observe the generation of spin entanglement between two particles through a Newtonian gravitational interaction~\cite{boseSpinEntanglementWitness2017,marlettoGravitationallyInducedEntanglement2017}. Motivated by earlier ideas~\cite{kafriClassicalChannelModel2014}, this approach is based upon the quantum information theoretic definition of quantum versus classical channels, which is elevated to a definition of ``quantumness'' of the gravitational field. The proposed experimental test consists of two particles in adjacent Stern-Gerlach interferometers, such that the mutual gravitational force between both particles results in observable entanglement.

As far as the distinction between perturbative quantum gravity and the semi-classical Einstein equations goes, these two experimental approaches are equivalent. In the nonrelativistic limit, perturbative quantum gravity results in a linear Schrödinger equation for two particles with a Newtonian potential
\begin{equation}\label{eqn:pot-linear}
 V = -\frac{G m_1 m_2}{\abs{\hat{\vec r}_1 - \hat{\vec r}_2}}\,,
\end{equation}
whereas the SN equation comprises the nonlinear two-particle potential
\begin{equation}\label{eqn:pot-sn}
V = -\frac{G m_1 m_2}{\abs{\erw{\hat{\vec r}_1} - \hat{\vec r}_2}} -\frac{G m_1 m_2}{\abs{\hat{\vec r}_1 - \erw{\hat{\vec r}_2}}} + \sum_\mathrm{particles} V_\mathrm{self}\,.
\end{equation}
Experimental confirmation of self-gravitational forces~\cite{yangMacroscopicQuantumMechanics2013,grossardtOptomechanicalTestSchrodingerNewton2016} would, therefore, definitively falsify the quantum potential~\eref{eqn:pot-linear}, while experimental evidence for entanglement would rule out the SN potential~\eref{eqn:pot-sn}. Only if one also considers alternative models which have neither of those potentials as a limit, the selectivity of both types of experiment differs.\footnote{The author is aware of only one physical model of this kind, namely that by Tilloy and Di\'osi~\cite{tilloySourcingSemiclassicalGravity2016} which suggests objective wave function collapse events as the source for spacetime curvature.}

Although there is justifiable hope for experimental evidence in the near future, the requirements regarding large masses, effective cooling, long decoherence times, among others, pose a challenge. Experimental proposals to detect gravitational spin entanglement~\cite{boseSpinEntanglementWitness2017} are strongly constrained by acceleration noise requirements~\cite{grossardtAccelerationNoiseConstraints2020}. On these grounds, it is important to consider a variety of ideas for experimental tests, in order to select the most promising scenario.

Here we discuss the possibility to adapt the experimental set-up of a Stern-Gerlach interferometer, as in reference~\cite{boseSpinEntanglementWitness2017}, for the purpose of testing the SN equation. Instead of the two adjacent interferometers for two particles, we consider a single particle in a single interferometer.
The usual treatment of weak potentials in terms of pure phase shifts is, however, not necessarily adequate, as it relies on the plane wave approximation, whereas the wave functions in realistic experimental situations are generally well localized. We, therefore, present a rigorous theoretical treatment with only well-justified approximations.

In \sref{sec:two} we define the basic model for gravitational self-interaction of a particle with spin in a superposition of two trajectories. \Sref{sec:gauss} discusses the limit of a wide Gaussian wave function, whereas we discuss the opposite case of a well localized wave function in \sref{sec:localized}. Finally, we discuss experimental consequences in \sref{sec:discussion}, concluding with a summary of our results in \sref{sec:conclusion}.

\section{Self-gravity in a Stern-Gerlach interferometer}\label{sec:two}

\begin{figure}
\centering
\includegraphics[scale=0.5]{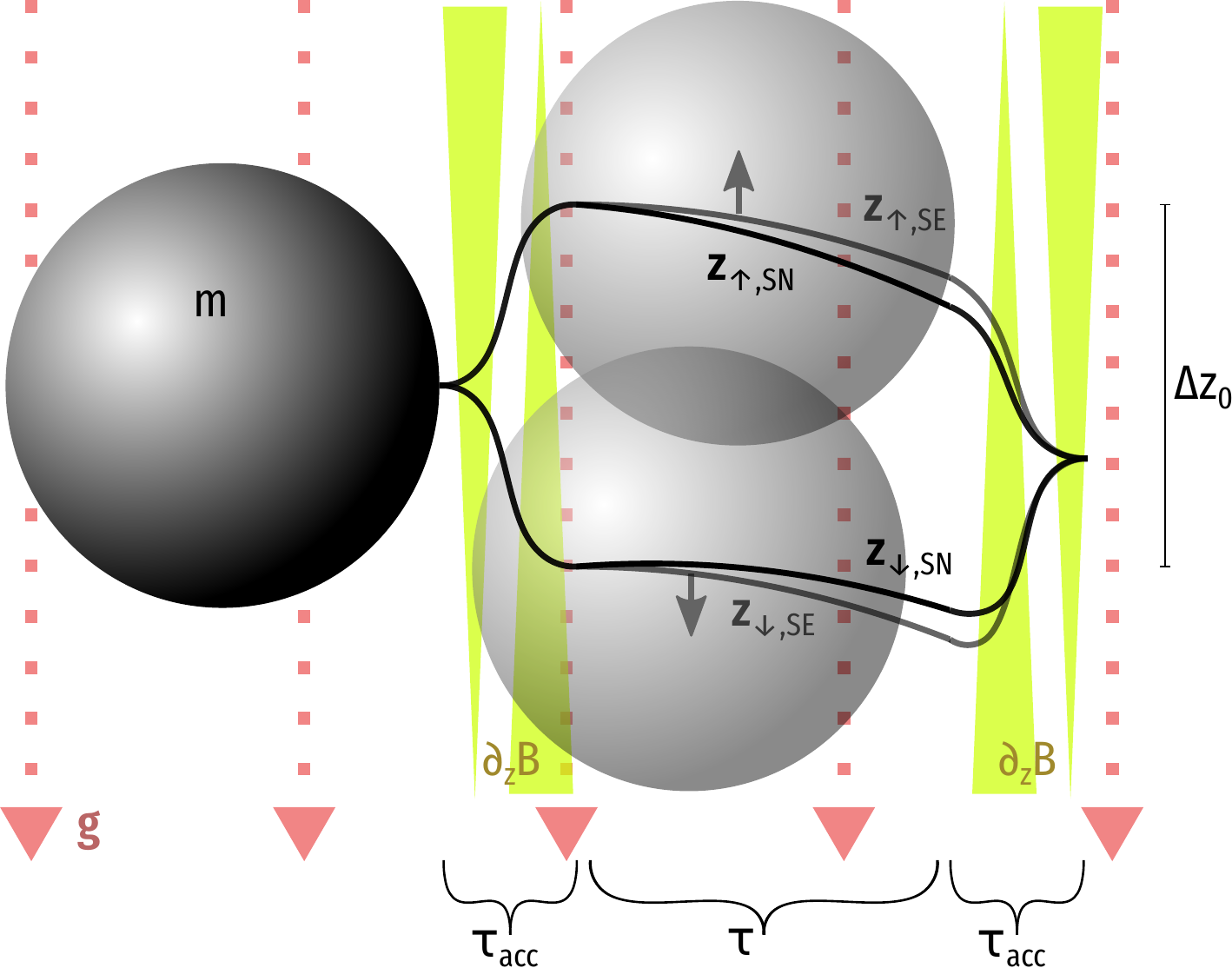}
 \caption{Schematic of the experimental set-up: a spin-$\frac12$\ particle of mass $m$ is brought into a spatial superposition state by being exposed to a magnetic field gradient $\partial_z B$ for a time $\taua$, with a spin-flip after $\taua/2$ resulting to deceleration and a free fall phase of time $\tau$. The trajectory taken under the self-gravitational influence of the SN equation slightly differs from the free trajectory, resulting in a different phase shift from the external acceleration $\vec g$.\label{fig:setup}}
\end{figure}

The situation we have in mind is a single spin-$\frac12$\ particle whose trajectory is split in two by a magnetic field gradient and reunited by an inverse field, allowing for phase dependent interference. \Fref{fig:setup} is an idealized depiction of the set-up, following the protocol by Bose et al.~\cite{boseSpinEntanglementWitness2017}, where the field gradient is effectively inverted by a spin flip or switched off by transferring the spin from the electrons to the nuclei. It also shows the intuitive expectation if self-gravity is involved: the spin-up and spin-down parts should attract each other and result in slightly different trajectories. As will become clear from the succeeding discussion, this intuition is, however, misleading.

The state of this particle is given by
\begin{equation}\label{eqn:spin-onehalf-state}
\ket{\Psi(t)} = 
\alpha \up \otimes \int \rmd^3 r \, \psi_\uparrow(t,\vec r) \ket{\vec r}
+ \beta \down \otimes \int \rmd^3 r \,  \psi_\downarrow(t,\vec r) \ket{\vec r} \,.
\end{equation}
with $\abs{\alpha}^2 + \abs{\beta}^2 = 1$ and $\psi_\uparrow(0,\vec r) = \psi_\downarrow(0,\vec r)$.

In the final state at time $T = \tau + 2 \taua$, we measure the spin in $x$-direction. We obtain the reduced density matrix for the spin after tracing out the position degrees of freedom in the state~\eref{eqn:spin-onehalf-state}:
\begin{equation}
\fl \hat{\rho} = \int \rmd^3 r \left(
\abs{\alpha}^2 \abs{\psi_\uparrow}^2 \up\bup
+ \alpha \beta^* \psi_\uparrow \psi_\downarrow^* \down\bup
+ \alpha^* \beta \psi_\uparrow^* \psi_\downarrow \up\bdown
+ \abs{\beta}^2 \abs{\psi_\downarrow}^2 \down\bdown \right) \,.
\end{equation}
The expectation value for the spin operator $\hat{\sigma}_x = \up\bdown + \down\bup$ is then obtained as
\begin{equation}\label{eqn:expectation-sigma-x}
\erw{\hat{\sigma}_x} = \Tr \hat{\rho} \hat{\sigma}_x
= \mathrm{Re} \left[2 \alpha \beta^* \int \rmd^3 r \, \psi_\uparrow \psi_\downarrow^* \right] \,.
\end{equation}
In an ideal interferometric scenario, there is a large overlap of the spatial wave functions, such that the final states differ only by a phase, $\psi_\uparrow(T,\vec r) \approx \rme^{\rmi \Delta \phi} \psi_\downarrow(T,\vec r)$, and we find $\erw{\hat{\sigma}_x} = \sqrt{1-\delta^2} \cos (\phi_{\alpha\beta} + \Delta \phi)$, i.\,e.\ constructive and destructive interference depending on the phase difference, where $\delta = \abs{\alpha}^2 - \abs{\beta}^2$ measures the asymmetry of the superposition, and $\phi_{\alpha\beta}$ is the relative phase between the parameters $\alpha$ and $\beta$.
If the spatial overlap of the final wave functions becomes smaller, visibility of interference is suppressed.

In the case of semi-classical gravity, the evolution of the wave function~\eref{eqn:spin-onehalf-state} is described by the Hamiltonian
\begin{equation}\label{eqn:full-hamiltonian}
 \hat{H} = \hat{I} \otimes \left(-\frac{\hbar^2}{2 m}\nabla^2 + V_\mathrm{ext} + \abs{\alpha}^2 U_\uparrow + \abs{\beta}^2 U_\downarrow\right) + \hat{\sigma}_z \otimes V_\mathrm{acc}
\end{equation}
where $\hat{\sigma}_z$ denotes the third Pauli matrix and $\hat{I}$ the identity in spin space. $V_\mathrm{acc}$ is a spin dependent potential responsible for the trajectory split which will be homogeneous for the further discussion and we include an external potential, which will be either negligible or a homogeneous acceleration, $V_\mathrm{ext} = m g z$, for the subsequent discussion. $U_\ud$ are the self-gravitational potentials corresponding to the spatial wave functions $\psi_\ud$ and take the explicit form~\cite{giuliniCentreofmassMotionMultiparticle2014,grossardtEffectsNewtonianGravitational2016}
\numparts
\begin{eqnarray}
 U_{\ud}(t,\vec r) &= \int \rmd^3 r' \abs{\psi_\ud(t,\vec r')}^2 I_\rho(\vec r - \vec r') \\
I_\rho(\vec d) &= -G \, \int \rmd^3 \vec r\, \rmd^3 \vec r' \,
\frac{\rho(\vec r)\,\rho(\vec r')}{\abs{\vec r - \vec r' + \vec d}} \,, \label{eqn:irho}
\end{eqnarray}
\endnumparts
where $\rho(\vec r)$ is the mass density distribution associated with the particle, and $I_\rho(\vec d)$ is a function of only the magnitude $d = \abs{\vec d}$ for spherically symmetric $\rho(\vec r)$.
If the particle is approximately point-like and, hence, its mass distribution a delta distribution, one recovers the single particle potential~\eref{eqn:pot-sn-self}. 

Crucial about the semi-classical potential in the Hamiltonian~\eref{eqn:full-hamiltonian} is that both spin states are subject to the same potential which depends on the \emph{full} state, i.\,e.\ the motion of the $\up$ part depends on the trajectory of the $\down$ part and vice versa.
Due to this state dependence of the potentials $U_\ud$, the Schrödinger equation
\begin{eqnarray}\label{eqn:schroedinger-inseparable}
0 &= \alpha \left(\rmi \hbar \partial_t + \frac{\hbar^2}{2 m}\nabla^2 - V_\mathrm{ext} - V_\mathrm{acc} - \abs{\alpha}^2 U_\uparrow - \abs{\beta}^2 U_\downarrow\right) \up \otimes \psi_\uparrow \nnl
&\bleq + \beta \left(\rmi \hbar \partial_t + \frac{\hbar^2}{2 m}\nabla^2 - V_\mathrm{ext} + V_\mathrm{acc} - \abs{\alpha}^2 U_\uparrow - \abs{\beta}^2 U_\downarrow\right) \down \otimes \psi_\downarrow
\end{eqnarray}
then becomes inseparable, which in turn induces a time dependence of the coefficients $\alpha$ and $\beta$.
Assuming that the gravitational potential $U_\ud$ is weak, however, we can take a perturbative approach, by first solving the separable equations for $U_\ud = 0$, i.\,e. finding solutions to
\begin{equation}\label{eqn:zero-order-se}
\rmi \hbar \partial_t \psi^{(0)}_\ud = \left(-\frac{\hbar^2}{2 m}\nabla^2 - \vec F_\ud(t) \cdot \vec r\right) \psi^{(0)}_\ud \,,
\end{equation}
and then solving the now also separable system at $n$-th order
\begin{equation}\label{eqn:se-n}
\rmi \hbar \partial_t \psi^{(n)}_\ud = \left(-\frac{\hbar^2}{2 m}\nabla^2 - \vec F_\ud(t) \cdot \vec r + \abs{\alpha}^2 U^{(n-1)}_\uparrow + \abs{\beta}^2 U^{(n-1)}_\downarrow\right) \psi^{(n)}_\ud \,,
\end{equation}
where $U^{(n-1)}_\ud$ are calculated from the spatial wave functions $\psi^{(n-1)}_\ud$ and we collated the homogeneous potentials $V_\mathrm{ext}$ and $V_\mathrm{acc}$ in the force $\vec F_\ud$.
For the specific acceleration sequence as depicted in \fref{fig:setup}, the external force is
\begin{equation}
\vec F_\ud(t) = -m \times \left\{ \begin{array}{ll}
\vec g \mp a \vec{e}_z
&\mbox{for}\ t \in [0,\frac{\taua}{2}] \cup [\tau+\frac{3\taua}{2},\tau+2\taua]\,,\\
\vec g \pm a \vec{e}_z
&\mbox{for}\ t \in [\frac{\taua}{2},\taua] \cup [\tau+\taua,\tau+\frac{3\taua}{2}]\,,\\
\vec g
&\mbox{everywhere else.}
\end{array}\right.
\end{equation}
The solution of \eref{eqn:se-n} is then given by (cf. \ref{app:theorem})
\begin{eqnarray}\label{eqn:psi-chi}
\psi^{(n)}_\ud(t,\vec r) &= \rme^{\rmi \varphi_\ud(t,\vec r)} \chi_\ud^{(n)}(t,\vec r-\vec u_\ud(t)) \\
\varphi_\ud(t,\vec r) &= \frac{m}{\hbar} \,\vec r \cdot \dot{\vec u}_\ud(t) - \frac{m}{2 \hbar} \int_0^t \rmd t' \dot{\vec u}_\ud(t')^2\,, \label{eqn:phasealpha}
\end{eqnarray}
where $\vec u_\ud(t)$ solves the equations of motion
\begin{equation}\label{eqn:eqmotion}
\ddot{\vec u}_\ud(t) = \frac{\vec F_\ud(t)}{m} 
\qquad\mbox{with}\qquad
\vec u_\ud(0)= \dot{\vec u}_\ud(0) = 0 \,,
\end{equation}
and $\chi_\ud^{(n)}(t,\vec r)$ solves the Schrödinger equation in the rest frame,
\begin{equation}\label{eqn:se-rest-frame}
 \rmi \hbar \dot{\chi}_\ud^{(n)}(t,\vec r) = -\frac{\hbar^2}{2m} \nabla^2 \chi_\ud^{(n)}(t,\vec r) 
 + U^{(n-1)}(t,\vec r + \vec u_\ud(t)) \chi_\ud(t,\vec r)\,,
\end{equation}
with the same initial conditions $\chi_\ud(0,\vec r) = \psi_\ud(0,\vec r)$ and the potentials
\begin{equation}
\fl U^{(n)}(t,\vec r) = \int \rmd^3 r' \,
\left(\abs{\alpha}^2 \abs{\chi_\uparrow^{(n)}(t,\vec r'-\vec u_\uparrow(t))}^2
+ \abs{\beta}^2 \abs{\chi_\downarrow^{(n)}(t,\vec r'-\vec u_\downarrow(t))}^2 
\right) I_\rho(\vec r - \vec r') \,.
\end{equation}
At order $n=0$ the solutions differ only through the classical trajectories $\vec u_\ud$ and their wave functions are identical, $\chi^{(0)}_\uparrow = \chi^{(0)}_\downarrow$. Hence, if we introduce
\begin{equation}\label{eqn:utilde}
 \widetilde{U}(t,\vec r) = \int \rmd^3 r' \,
\abs{\chi^{(0)}(t,\vec r')}^2 I_\rho(\vec r - \vec r') \,,
\end{equation}
we have
\begin{equation}
U^{(0)}(t,\vec r) =
\abs{\alpha}^2 \widetilde{U}(t,\vec r - \vec u_\uparrow(t))
+ \abs{\beta}^2 \widetilde{U}(t,\vec r - \vec u_\downarrow(t))\,,
\end{equation}
and the zeroth and first order Schrödinger equations to be solved are
\numparts
\begin{eqnarray}
\fl \rmi \hbar \dot{\chi}^{(0)}(t,\vec r) &= -\frac{\hbar^2}{2m} \nabla^2 \chi^{(0)}(t,\vec r)\,, \\
\fl \rmi \hbar \dot{\chi}_\uparrow^{(1)}(t,\vec r) &= \left(-\frac{\hbar^2}{2m} \nabla^2
+ \abs{\alpha}^2 \widetilde{U}(t,\vec r)
+ \abs{\beta}^2 \widetilde{U}(t,\vec r + \Delta \vec u(t)) \right) \chi_\uparrow^{(1)}(t,\vec r)\,, \label{eqn:first-order-se-up} \\
\fl \rmi \hbar \dot{\chi}_\downarrow^{(1)}(t,\vec r) &= \left(-\frac{\hbar^2}{2m} \nabla^2
+ \abs{\alpha}^2 \widetilde{U}(t,\vec r - \Delta \vec u(t))
+ \abs{\beta}^2 \widetilde{U}(t,\vec r) \right) \chi_\downarrow^{(1)}(t,\vec r)\,.\label{eqn:first-order-se-down}
\end{eqnarray}
\endnumparts
The difference $\Delta \vec u = \vec u_\uparrow - \vec u_\downarrow$ can be straightforwardly obtained from the equations of motion \eref{eqn:eqmotion} and is independent of the external force $\vec g$. For the specific acceleration sequence as in \fref{fig:setup} it solves
\begin{equation}
\Delta \ddot{u} = \left\{ \begin{array}{ll}
2 a
&\mbox{for}\ t \in [0,\frac{\taua}{2}] \cup [\tau+\frac{3\taua}{2},\tau+2\taua]\,,\\
- 2 a
&\mbox{for}\ t \in [\frac{\taua}{2},\taua] \cup [\tau+\taua,\tau+\frac{3\taua}{2}]\,,\\
0
&\mbox{everywhere else,}
\end{array}\right.\label{eqn:uddot}
\end{equation}
with $\Delta u(0) = \Delta \dot{u}(0) = 0$. It also follows immediately that for the sum $u_\uparrow + u_\downarrow = -gt^2$.
The phase \eref{eqn:phasealpha} will generally depend on the external force. For $\vec g$ in $z$-direction, we find
\begin{eqnarray}
 \Delta \varphi(t,z) &= \varphi_\uparrow - \varphi_\downarrow 
 = \frac{m}{\hbar} \left( z \Delta \dot{u}(t)
 + g \int_0^t \rmd t' \, t' \Delta \dot{u}(t') \right) \\
 &= \frac{m}{\hbar} \left( \left(z + \frac{g t^2}{2}\right) \Delta \dot{u}(t)
 - \frac{g}{2} \int_0^t \rmd t' \, {t'}^2 \Delta \ddot{u}(t') \right)\,.
\end{eqnarray}
Note that this phase difference depends solely on the classical trajectories. Self-gravity effects manifest themselves only in the evolution of the wave functions $\chi_\ud$.
Excluding the first term proportional to $\Delta \dot{u}$, one finds the global phase
\begin{equation}\label{eqn:phig}
\varphi_g(t) 
 = - \frac{m g}{2 \hbar} \int_0^t \rmd t' \, {t'}^2 \Delta \ddot{u}(t')
 \stackrel{t=T}{=} -\frac{m a g}{2 \hbar} \taua^2 (\tau + \taua) \,.
\end{equation}
where the second equality is the final phase at the end of the sequence as in \fref{fig:setup} after time $T = \tau + 2 \taua$, which reproduces the well known result~\cite{colellaObservationGravitationallyInduced1975}.
The spin expectation value according to equation \eref{eqn:expectation-sigma-x} is then obtained as
\begin{eqnarray}\label{eqn:expectation-sigma-x-chi}
\erw{\hat{\sigma}_x}^{(n)}(t) &= \sqrt{1-\delta^2} \, \mathrm{Re} \Bigg[\rme^{\rmi \phi_{\alpha\beta}} \int \rmd z \, \rme^{\rmi \, \Delta \varphi(t,z)} 
\nnl &\bleq 
\chi^{(n)}_\uparrow\left(t,z+\frac{g t^2}{2}-\frac{\Delta u}{2}\right) {\chi^{(n)}_\downarrow}^*\left(t,z+\frac{g t^2}{2}+\frac{\Delta u}{2}\right) \Bigg] \nnl
&= \sqrt{1-\delta^2} \, \mathrm{Re} \Bigg[\rme^{\rmi (\phi_{\alpha\beta} + \varphi_g(t))} \int \rmd z \,
\exp\left(\frac{\rmi \,m}{2 \hbar} (2 z + \Delta u) \Delta \dot{u})\right)
\nnl &\bleq 
\chi^{(n)}_\uparrow\left(t, z\right) {\chi^{(n)}_\downarrow}^*\left(t, z+\Delta u\right) \Bigg]
\end{eqnarray}
where $\phi_{\alpha\beta}$ is the relative phase between the coefficients $\alpha$ and $\beta$.

For the discussion of self-gravitational corrections, it is advantageous to consider two different cases separately: first, the situation of a point-like particle in which the spreading of the wave function is large compared to the particle size; second, the opposite situation where the particle is well localized.

\section{Point-like particle with Gaussian wave function}\label{sec:gauss}

We assume spherically symmetric, Gaussian initial conditions. Then we find at the lowest order $n=0$ the free solutions with probability density
\begin{equation}
\abs{\chi^{(0)}(t,\vec r)}^2 = \left(2 \pi A^{(0)}(t)\right)^{-3/2}
\exp\left(-\frac{r^2}{2 A^{(0)}(t)}\right)
\end{equation}
with $A^{(0)}(t) = A_0 + \hbar^2 t^2/(4 m^2 A_0)$, and the potential
\numparts
\begin{eqnarray}\label{eqn:self-grav-pot-gauss}
\widetilde{U}(t,r) &= \left(2 \pi A^{(0)}(t)\right)^{-3/2} \int \rmd^3 r' \,
\exp\left(-\frac{{r'}^2}{2 A^{(0)}(t)}\right) I_\rho(\vec r - \vec r') \\
&\approx -\frac{G m^2}{r} \mathrm{erf}\left(\sqrt{\frac{r^2}{2 A^{(0)}(t)}}\right) \,.\label{eqn:self-grav-pot-gauss-approx}
\end{eqnarray}
\endnumparts
The potential~\eref{eqn:self-grav-pot-gauss} can be evaluated analytically and its derivation is presented in \ref{app:potential}. It is spherically symmetric and obtains its time dependence solely through the time dependence of $A^{(0)}(t)$. Equation \eref{eqn:self-grav-pot-gauss-approx} assumes a small particle radius, $R \ll \sqrt{A^{(0)}(t)}$.

In order to proceed further, we make a second approximation~\cite{colinCrucialTestsMacrorealist2016,grossardtApproximationsFreeEvolution2016}, assuming that---despite the nonseparability of the SN equation---the wave function remains a separable Gaussian,
\begin{equation} \label{eqn:wave-function-coherent}
\fl \chi^{(1)}_\ud(t,\vec r)
\approx \Upsilon(t,x,y) \left(2\pi A_\ud\right)^{-1/4}
\exp\left[-\frac{\left(z-\erw{z}_\ud\right)^2}{4 A_\ud}
\left(1 - \frac{\rmi B_\ud}{\hbar}\right)
+ \rmi \frac{\erw{p}_\ud z + f_\ud}{\hbar} \right] \,.
\end{equation}
Throughout this article we write $p$ for the momentum in $z$-direction, and the second moments are defined as
\numparts
\begin{eqnarray}
A_\ud(t) &= \erw{z^2}_\ud - \erw{z}_\ud^2 \\
B_\ud(t) &= \erw{z p + p z}_\ud - 2 \erw{z}_\ud \erw{p}_\ud \\
C_\ud(t) &= \erw{p^2}_\ud - \erw{p}_\ud^2 \,,
\end{eqnarray}
\endnumparts
which satisfy at any time $4 A_\ud C_\ud - B_\ud^2 = \hbar^2$. Furthermore, we introduce the phase
\begin{eqnarray}\label{eqn:phase-f}
f_\ud(t) &= f_0 - \frac{\erw{z}_\ud \erw{p}_\ud}{2} - \frac{\hbar^2}{4m} \int_0^t \frac{\rmd t'}{A_\ud(t')} \,,
\end{eqnarray}
chosen such that the wave function \eref{eqn:wave-function-coherent} solves the Schrödinger equation in the absence of the gravitational potential. 
The first and second moments satisfy the equations of motion
\numparts
\begin{eqnarray}
\partial_t \erw{z}_\ud &= \frac{\erw{p}_\ud}{m} \label{eqn:moments-z} \\
\partial_t \erw{p}_\ud &= -\erw{\partial_z V_\ud} \\
\partial_t A_\ud &= \frac{B_\ud}{m} \\
\partial_t B_\ud &= \frac{2 C_\ud}{m} - 2 \erw{(z-\erw{z}_\ud) (\partial_z V_\ud)} \\
\partial_t C_\ud &= - \erw{(p-\erw{p}_\ud) (\partial_z V_\ud) + (\partial_z V_\ud) (p-\erw{p}_\ud)} \label{eqn:moments-c} \,,
\end{eqnarray}
\endnumparts
where the potential $V_\ud$ is the one from equations \eref{eqn:first-order-se-up} and \eref{eqn:first-order-se-down}.

In principle, all expectation values need to be evaluated with the wave function $\chi_\ud^{(1)}$ at first order. However, the expectation values of the gravitational potential are already of higher order and can, therefore, be evaluated using the zeroth order wave function $\chi^{(0)}$ at the same level of approximation.
The potential $\widetilde{U}$ has the symmetries
\numparts
\begin{eqnarray}
\erw{\widetilde{U}(t,\vec r-\Delta \vec u} &= \erw{\widetilde{U}(t,\vec r+\Delta \vec u} \\
\erw{\partial_z \widetilde{U}(t,\vec r-\Delta \vec u} &= -\erw{\partial_z \widetilde{U}(t,\vec r+\Delta \vec u} \\
\erw{z \partial_z \widetilde{U}(t,\vec r-\Delta \vec u} &= \erw{z \partial_z \widetilde{U}(t,\vec r+\Delta \vec u} \,.
\end{eqnarray}
\endnumparts
With the results from \ref{app:expvals}, noting that $\erw{z}_\ud = \erw{p}_\ud = 0$ if evaluated with the zeroth order wave function $\chi^{(0)}$, and writing again $\delta = \abs{\alpha}^2 - \abs{\beta}^2$, we have
\numparts
\begin{eqnarray}
\fl \partial_t \erw{p}_\ud &= \frac{1}{2} (\delta \mp 1)\erw{\partial_z \widetilde{U}(t,\vec r + \Delta \vec u)}_0
\nnl\fl &= \frac{G m^2}{2 \Delta u^2} (\delta \mp 1)  \left(  \exp\left(-\frac{\Delta u^2}{4A^{(0)}}\right)\frac{\Delta u}{\sqrt{\pi A^{(0)}}}  -I_0^-\left(\frac{\Delta u}{\sqrt{2A^{(0)}}}\right) \right)
\label{eqn:expval-pz} \\
\fl \partial_t B_\ud &= \frac{2 C_\ud}{m} 
- (1\pm\delta) \erw{z \partial_z \widetilde{U}(t,\vec r)}_0
- (1\mp\delta) \erw{z \partial_z \widetilde{U}(t,\vec r + \Delta \vec u)}_0 
\nnl\fl &= \frac{2 C_\ud}{m} 
+ \frac{G m^2}{6 \sqrt{\pi A^{(0)}}} \Bigg[ 1 \pm \delta + (1\mp \delta) \Bigg(
\left(3 + \frac{12 A^{(0)}}{\Delta u^2}\right) \exp\left(-\frac{\Delta u^2}{4A^{(0)}}\right) \nnl\fl &\bleq
- \frac{12 \sqrt{\pi {A^{(0)}}^3}}{\Delta u^3} I_0^-\left(\frac{\Delta u}{\sqrt{2A^{(0)}}}\right)\Bigg)\Bigg]
\label{eqn:expval-B}\\
\fl \partial_t C_\ud &= - \frac{1\pm\delta}{2} m \partial_t \erw{\widetilde{U}(t,\vec r)}_0
- \frac{1\mp\delta}{2} m \partial_t \erw{\widetilde{U}(t,\vec r - \Delta \vec u)}_0
\nnl\fl &= \frac{G m^3}{2} \partial_t  \left( \frac{1 \pm \delta}{\sqrt{\pi A^{(0)}}}
+ \frac{1 \mp \delta}{\Delta u} I_0^-\left(\frac{\Delta u}{\sqrt{2A^{(0)}}}\right)
\right)
\,,\label{eqn:expval-C}
\end{eqnarray}
\endnumparts
with the integral 
\begin{equation}
I_0^-(\sigma) = \int_0^\infty \rmd \rho \, \frac{\rme^{-\rho^2}}{\sqrt{\pi}} \left(\mathrm{erf}(\rho + \sigma) - \mathrm{erf}(\rho - \sigma)\right)
\end{equation}
as defined in \ref{app:expvals}.
Given the initial conditions $\erw{z}_\ud(0) = \erw{p}_\ud(0) = 0$, $A_\ud(0) = A_0$, as well as
$\partial_t A_\ud \vert_{t=0} = 0$
we must then solve
\numparts
\begin{eqnarray}
\fl \partial_t^2 \erw{z}_\ud &= -\frac{G m}{2 A^{(0)}} (\delta \mp 1)  F_1\left(\frac{\Delta u}{\sqrt{2A^{(0)}}}\right)
 \\
\fl \partial_t^2 A_\ud &= \frac{\hbar^2}{2 m^2 A_0} 
- \frac{2 G m}{\sqrt{\pi A_0}} 
+ \frac{7 G m}{6 \sqrt{\pi A^{(0)}}} \left[ (1\pm\delta)
 + (1\mp\delta) F_2\left(\frac{\Delta u}{\sqrt{2A^{(0)}}}\right) \right]
\label{eqn:diff-eq-A}
\end{eqnarray}
with
\begin{eqnarray}
F_1(\sigma) &= \frac{I_0^-(\sigma)}{2 \sigma^2} - \frac{\rme^{-\sigma^2/2}}{\sqrt{2\pi} \sigma} \\
F_2(\sigma) &= \frac{6}{14 \sigma^3} \left((\sigma^3 + 2 \sigma) \rme^{-\sigma^2/2}
+ (\sigma^2 - 1) \sqrt{2\pi} I_0^-(\sigma)\right) \,.
\end{eqnarray}
\endnumparts
Approximations to the functions $F_1$ and $F_2$ can be found in \ref{app:expvals}.

\subsection{Symmetric case} 
Let us first focus on the case $\delta = 0$. Defining $\Delta z = \erw{z}_\uparrow - \erw{z}_\downarrow$, $\Delta p = \erw{p}_\uparrow - \erw{p}_\downarrow$ we have
\begin{equation}\label{eqn:dt2-deltaz}
\partial_t^2 \Delta z = \frac{G m}{A^{(0)}} \,F_1\left(\frac{\Delta u}{\sqrt{2 A^{(0)}}}\right) \,, \qquad
\Delta p = m \partial_t \Delta z \,,
\end{equation}
and we find that in this case the behavior of the second moments is identical for the spin up and down components: $A_\ud \equiv A$, $B_\ud \equiv B$, $C_\ud \equiv C$, where $A$ solves \eref{eqn:diff-eq-A} and
\begin{equation}
 B = m \dot{A} \,, \qquad
 C = \frac{\hbar^2 + m^2 \dot{A}^2}{4 A} \,.
\end{equation}
We can then evaluate the integral in equation \eref{eqn:expectation-sigma-x-chi} and find
\numparts
\begin{equation}
 \erw{\hat{\sigma}_x} = \rme^{-\gamma_\mathrm{SN}} \cos (\phi_{\alpha\beta} + \Delta \phi)
\end{equation}
with
\begin{eqnarray}
\fl \gamma_\mathrm{SN} &= \frac{C (\Delta u + \Delta z)^2 - B (\Delta u + \Delta z)(m \,\Delta \dot{u} + \Delta p) + A (m \,\Delta \dot{u} + \Delta p)^2}{2 \hbar^2} \label{eqn:gamma-wide-sym} \\
\fl \Delta \phi &= -\frac{m g t^2}{2 \hbar} \Delta \dot{u} + \frac{m g}{\hbar} \int_0^t \rmd t' \, t' \Delta\dot{u}(t') \,.
\end{eqnarray}
\endnumparts
At the final time $T = \tau + 2 \taua$ we have $\Delta u = \Delta \dot{u} = 0$ and, thus,
\begin{equation}
 \erw{\hat{\sigma}_x}_f = \rme^{-\gamma_\mathrm{SN}} 
 \cos \left(\phi_{\alpha\beta} + \varphi_g \right) \,,
\end{equation}
which yields the usual oscillation $\varphi_g$ without any self-gravitational effect on the phase. The additional exponential dampening factor suppresses the visibility of the interference pattern. Note that $\Delta z$ as well as $\Delta p$ are self-gravitational corrections of order $G$, implying that the dampening factor is of quadratic order in $G$. We can, thus, approximate $A$, $B$, and $C$ by their unperturbed values and find
\begin{equation}\label{eqn:gamma-quadratic-order}
 \gamma_\mathrm{SN}^0 = \frac{1}{8 A_0} \left( \Delta z^2
+ 4 A_0^2 \frac{\Delta p^2}{\hbar^2}
- 2 \Delta z \frac{\Delta p}{m} T
+ \frac{\Delta p^2}{m^2} T^2
\right) \,.
\end{equation}
It must be remarked that the expansion chosen at the beginning with the Schrödinger equation \eref{eqn:se-n} technically implies that all results here, which are for $n=1$, are only accurate to linear order in $G$. However, it is easy to see that, due to the occurrence of the quadratic terms of $\Delta z$ and $\Delta p$, the dampening factor $\gamma_\mathrm{SN}$ will always be of one order higher than the phase. At the next order $n=2$ we would find corrections to the phase quadratic in $G$ and cubic corrections to $\gamma_\mathrm{SN}$. Therefore, we may use the result \eref{eqn:gamma-quadratic-order} as an appropriate approximation, nonetheless.

Consider the concrete situation in which the particle of radius $R$ is split to its maximum separation $\Delta u_\mathrm{max} = a \taua^2 / 2$ in a short time $\taua \ll \tau$, such that $\Delta u_\mathrm{max} \gg \sqrt{A_0} \gg R$. We then can approximately assume constant separation throughout the experiment and neglect the spreading of the wave function, i.\,e. take $A^{(0)} \approx A_0$ constant. One finds $\Delta z \approx G m t^2/(4 \Delta u_\mathrm{max})$ and the second term in equation \eref{eqn:gamma-quadratic-order} dominates:
\begin{equation}\label{eqn:gamma0-wide-approx}
 \gamma_\mathrm{SN}^0 \approx \frac{G^2 m^4 \tau^2 A_0}{8 \hbar^2 \Delta u_\mathrm{max}^4}
 \ll \frac{G m^4 \tau^2}{8 \hbar^2 R^2} \,.
\end{equation}

\subsection{Asymmetrical case} 
For the general case $\delta \not= 0$, first note that
\begin{equation}
 \erw{z}_\ud = (-\delta \pm 1) \frac{\Delta z}{2} \qquad\mbox{and}\qquad 
 \erw{p}_\ud = (-\delta \pm 1) \frac{\Delta p}{2}
\end{equation}
with the solutions of equation \eref{eqn:dt2-deltaz} unaltered. We can further write
\begin{equation}\label{eqn:Aud-to-Apm}
A_\ud = A^{(0)} + A_+ \pm \delta A_- \,,
\end{equation}
where $A_\pm$ are of order $G$. Hence, calculating the expectation value \eref{eqn:expectation-sigma-x-chi} from the wave function \eref{eqn:wave-function-coherent}, and considering terms up to linear order in $G$ for the phase and of quadratic order for the dampening, we find at the final time $T = \tau + 2 \taua$:
\numparts\label{eqnsys:spin-expectation-asymmetrical}
\begin{eqnarray}\label{eqn:spin-expectation-asymmetrical}
 \fl \erw{\hat{\sigma}_x} &= \sqrt{1-\delta^2} \rme^{-\gamma_\mathrm{SN}} 
 \cos \left(\phi_{\alpha\beta} + \varphi_g + \Delta\phi_\mathrm{SN} \right) \\
 \fl \Delta\phi_\mathrm{SN} &= \frac{\delta \, m}{2 \hbar} \left( \dot{A}_- - \frac{\dot{A}^{(0)} A_-}{A^{(0)}} \right)_{t=T}
 + \frac{\delta \, \hbar}{2 m} \int_0^T \rmd t \, \frac{A_-(t)}{{A^{(0)}(t)}^2} \label{eqn:phase-asymmetrical}\\
 \fl \gamma_\mathrm{SN} &= \gamma_\mathrm{SN}^0 + \delta^2 m^2 \frac{4 \hbar^2 A_-^2 + 4 m^2 A_0^2 \dot{A}_-^2 - 4 \hbar^2 A_- \dot{A}_- T + \hbar^2 \dot{A}_-^2 T^2}{4 \hbar^2 (4 m^2 A_0^2 + \hbar^2 T^2)} \,.\label{eqn:gamma-asymmetrical}
\end{eqnarray}
\endnumparts
The integral in equation \eref{eqn:phase-asymmetrical} is easier evaluated after threefold partial integration, as demonstrated in \ref{app:fint}.

Now we introduce the frequency $\omega = \hbar/(2 m A_0)$ and express the acceleration $a$ in terms of the frequency $\Omega^4 = a^2/(2A_0)$, such that
with the solutions to \eref{eqn:uddot} one finds
\begin{eqnarray}\label{eqn:delta-u-solution}
\fl \eta(t) &= \frac{\Delta u}{\sqrt{2A_0(1+\omega^2 t^2)}} \nnl
\fl &= \frac{\Omega^2}{\sqrt{1+\omega^2 t^2}} \left\{ \begin{array}{ll}
t^2
&\mbox{for}\ t \in [0,\frac{\taua}{2}] \,,\\
t^2 - \frac{1}{2} (2 t - \taua)^2
&\mbox{for}\ t \in [\frac{\taua}{2},\taua] \,,\\
\frac{1}{2} \taua^2
&\mbox{for}\ t \in [\taua,\tau+\taua] \,,\\
\frac{1}{2} \taua^2 - (t - \tau - \taua)^2
&\mbox{for}\ t \in [\tau+\taua,\tau+\frac{3\taua}{2}] \,,\\
(t - \tau - 2\taua)^2
&\mbox{for}\ t \in [\tau+\frac{3\taua}{2},\tau+2\taua] \,.
\end{array}\right.
\end{eqnarray}
Then $A_\pm$ are the solution of
\numparts
\begin{eqnarray}
\partial_t^2 A_+ &=  \frac{7 G m}{6 \sqrt{\pi A_0}} \times 
\frac{1
 + F_2\left(\eta(t)\right)}{\sqrt{1 + \omega^2 t^2}}  - \frac{2 G m}{\sqrt{\pi A_0}} \label{eqn:dt-aepsilon}\\
\partial_t^2 A_- &= \frac{7 G m}{6 \sqrt{\pi A_0}} \times 
\frac{1
 - F_2\left(\eta(t)\right)}{\sqrt{1 + \omega^2 t^2}} \label{eqn:dt-adelta} \,,
\end{eqnarray}
\endnumparts
with $A_\pm$ and their time derivatives vanishing at $t=0$. The function $F_2$ vanishes for the sum of both equations which, therefore, can be integrated to yield
\begin{eqnarray}
 A_+ + A_- &= \frac{G m}{\omega^2 \sqrt{\pi A_0}} \left( \frac{7}{3} - \omega^2 t^2 - \frac{7}{3} \sqrt{1+ \omega^2 t^2} + \frac{7 \omega t}{3} \mathrm{arcsinh}(\omega t) \right)\,.
\end{eqnarray}
Up to linear order in $G$ one finds
the phase shift 
\begin{equation}
\Delta \phi_\mathrm{SN} = \delta \frac{m \omega}{2 \hbar} \left[ \frac{\dot{A}_-(T)}{\omega} - \frac{2 \omega T A_-(T)}{1+\omega^2 T^2}
+  \int_0^T \rmd t \, \frac{4 \omega \, A_-(t)}{(1 + \omega^2 t^2)^2} \right]\,.
\label{eqn:phase-gauss}
\end{equation}
The wave function width $A_0$ became absorbed in the frequency $\omega$. The expression in square brackets can be found through numerical integration and depends only on the two frequency values $\omega$ and $\Omega$, as well as the times $\tau$ and $\taua$. In order to determine the phase difference, the particle mass enters as a third parameter through the dimensionless prefactor.

In the experimental situations of interest, the spin dependent acceleration results from a magnetic field gradient, i.\,e. $m a = \mu_B \partial_z B$ with the Bohr magneton $\mu_B$. Assuming a spherical particle of radius $R$ and mass density $\rho$, and taking into account that we are in the limit where $R^2 \ll A_0$, we find that
\begin{equation}
\fl \varepsilon = \frac{\omega}{\Omega} \ll \sqrt{\frac{3 \hbar^2}{\sqrt{2} \,8 \pi \,\rho \,\mu_B \,\partial_z B \, R^6}}
\approx 10^{-7} \,\left(\frac{\rho}{\gram\per\centi\meter\cubed}\right)^{-1/2}
\,\left(\frac{\partial_z B}{\tesla\per\meter}\right)^{-1/2}
\,\left(\frac{R}{\micro\meter}\right)^{-3} \,.
\end{equation}
Hence, for realistic parameters where self-gravity plays a role, we will usually have $\varepsilon \ll 1$. Introducing the dimensionless time parameter $\theta = \Omega t$, and assuming also $\omega t \ll 1$, we can approximate the phase \eref{eqn:phase-gauss} by
\begin{equation}
\Delta \phi_\mathrm{SN} \approx  G \delta \sqrt{\frac{m^5}{2 \pi \hbar^3 \omega}} \widetilde{\phi}
\end{equation}
where, defining the dimensionless $A_\delta = \omega^2 \sqrt{\pi A_0} A_- / (G m)$, $\widetilde{\phi}$ solves
\numparts
\begin{eqnarray}
\partial_\theta \widetilde{\phi} &= \frac{\partial_\theta^2 A_\delta(\theta)}{\varepsilon} - 2 \varepsilon \theta \partial_\theta A_\delta(\theta) + 2 \varepsilon A_\delta(\theta) \\
\partial_\theta^2 A_\delta(\theta) &= \frac{7 \varepsilon^2}{6} (1-F_2(\eta)) \,.
\end{eqnarray}
\endnumparts
We have $F_2(\eta) \approx 1$ for small $\eta$ but $F_2 \to 0$ like $1/\eta$ as $\eta \gg 1$. Assuming that $\taua \gg 2/\Omega$ we can then, in a rather crude approximation, neglect $F_2$ entirely, finding $A_\delta \approx 7 \varepsilon^2 \theta^2 / 12$ and, in conclusion, $\widetilde\phi \approx 7 \varepsilon \theta^3 / 36$. Reinserting everything, we find 
\begin{equation}\label{eqn:approx-wide-phase}
\Delta \phi_\mathrm{SN} \approx \frac{\delta \, G m^2 \, a \, T^3}{2 \sqrt{2\pi} \,\hbar\, A_0} \,.
\end{equation}

\section{Localized particle}\label{sec:localized}

In the previous section we considered a Gaussian wave function much wider than the particle radius $R$. In the opposite case of a well localized particle, regardless of the shape of the wave function, we can approximate the potential \eref{eqn:utilde} by Taylor expansion around $\vec r - \vec r' = \vec 0$ to quadratic order~\cite{yangMacroscopicQuantumMechanics2013,giuliniCentreofmassMotionMultiparticle2014}.
We must, however, take into account that in equations \eref{eqn:first-order-se-up} and \eref{eqn:first-order-se-down} we also have the occurrence of $\widetilde{U}(t,\vec r \pm \Delta \vec u)$ with the offset $\pm \Delta \vec u$ acting on the wave function at $\vec r$. For a generic offset $\vec s = s \vec e_z$ in $z$-direction and for a spherically symmetric mass distribution $\rho(\vec r)$, we can approximate:
\begin{eqnarray}\label{eqn:u-quadratic}
\widetilde{U}^{(n)}_\ud(t,\vec r + \vec s) &= 
\int \rmd^3 r' \, \abs{\chi_\ud^{(n)}(t,\vec r')}^2 I_\rho(\abs{\vec r + \vec s - \vec r'}) \nnl
&= I_\rho(s) + I_\rho'(s) \left(z-\erw{z}^{(n)}_\ud\right) \nnl
&\bleq + \frac{I_\rho'(s)}{2s} \left(x^2 + \erw{x^2}^{(n)} 
+ y^2 + \erw{y^2}^{(n)} \right) \nnl
&\bleq + \frac{I_\rho''(s)}{2} \left(z^2-2 z \erw{z}^{(n)}_\ud + \erw{z^2}^{(n)}_\ud  \right) \,,
\end{eqnarray}
where we assumed $\erw{x}^{(n)} = \erw{y}^{(n)} = 0$ without loss of generality. Note that $I_\rho$ and $I_\rho''$ are symmetric, whereas $I_\rho'$ is antisymmetric. In the limit $s \to 0$ we find $I_\rho'(0) = 0$, $I_\rho'(s)/s \to I_\rho''(0)$ and, therefore, the known result
\begin{equation}\label{eqn:u-quadratic-origin}
\widetilde{U}^{(n)}_\ud(t,\vec r) = I_\rho(0) + \frac{I_\rho''(0)}{2} \left(\vec r^2 + \erw{\vec r^2}^{(n)}_\ud -2 z \erw{z}^{(n)}_\ud \right) \,.
\end{equation}

We begin with noticing that the potential \eref{eqn:u-quadratic} is exactly separable, and that the considerations in equations \eref{eqn:utilde}--\eref{eqn:first-order-se-down} can be generalized to higher orders $n$ to yield the Schrödinger equations
\numparts
\begin{equation}\label{eqn:schroedinger-narrow-n}
 \rmi \hbar \dot{\chi}_\ud^{(n)}(t,z) = \left(-\frac{\hbar^2}{2m} \partial_z^2
+ V^{(n-1)}_\ud(t,z) \right) \chi_\uparrow^{(n)}(t,z)
\end{equation}
with the potentials
\begin{eqnarray}
V^{(n)}_\uparrow(t,z) &= \abs{\alpha}^2 \widetilde{U}^{(n)}_\uparrow(t,\vec r) + \abs{\beta}^2 \widetilde{U}^{(n)}_\downarrow(t,\vec r + \Delta \vec u) \\ 
V^{(n)}_\downarrow(t,z) &= \abs{\alpha}^2 \widetilde{U}^{(n)}_\uparrow(t,\vec r -  \Delta \vec u) + \abs{\beta}^2 \widetilde{U}^{(n)}_\downarrow(t,\vec r)
\end{eqnarray}
\endnumparts
Defining $\Delta z^{(n)} = \erw{z}^{(n)}_\uparrow - \erw{z}^{(n)}_\downarrow$, as well as $k_\pm(\Delta u) = I_\rho''(0) \pm I_\rho''(\Delta u)$, we can then follow the derivation \eref{eqn:moments-z}--\eref{eqn:moments-c} of the equations of motion for the first and second moments which take the much simpler form
\numparts
\begin{eqnarray}
\partial_t \erw{z}^{(n)}_\ud &= \frac{\erw{p}^{(n)}_\ud}{m} \\
\partial_t \erw{p}^{(n)}_\ud &= \frac{\delta \mp 1}{2} \left(I_\rho'(\Delta u) + I_\rho''(\Delta u) \Delta z^{(n-1)} \right)  \\
\partial_t A^{(n)}_\ud &= \frac{B^{(n)}_\ud}{m} \\
\partial_t B^{(n)}_\ud &= \frac{2 C^{(n)}_\ud}{m} - \left(k_+(\Delta u)
\pm \delta \,k_-(\Delta u)\right) A^{(n-1)}_\ud \\
\partial_t C^{(n)}_\ud &= -\frac{1}{2} \left(k_+(\Delta u)
\pm \delta \,k_-(\Delta u)\right) B^{(n-1)}_\ud \,,
\end{eqnarray}
\endnumparts
We can express the first moments through the single equation
\begin{equation}\label{eqn:eom-deltaz}
\partial_t^2 \Delta {z}^{(n)} = -\frac{1}{m} \left(I_\rho'(\Delta u) + I_\rho''(\Delta u) \Delta z^{(n-1)} \right) \,,
\end{equation}
together with the sum
$\erw{z}^{(n)}_\uparrow + \erw{z}^{(n)}_\downarrow = -\delta \,\Delta z^{(n)}$.

As before, in the symmetric case $\delta = 0$ we find $A^{(n)}_\ud = A^{(n)}$ which satisfy
\begin{equation}
 \partial_t^3 A^{(n)} = -\frac{1}{m} \left(2 k_+(\Delta u) \dot{A}^{(n-1)} + \Delta \dot{u} \, k_+'(\Delta u) A^{(n-1)} \right) \,.
\end{equation}
The Schrödinger equations~\eref{eqn:schroedinger-narrow-n} are then solved by the squeezed coherent states
\begin{equation}\label{eqn:squeezed-chi-narrow}
\fl \chi^{(n)}_\ud(t,z) = \left(2\pi A^{(n)}\right)^{-1/4}
\exp\left[-\frac{\left(z-\erw{z}^{(n)}_\ud\right)^2}{4 A^{(n)}}
\left(1 - \frac{\rmi}{\hbar} B^{(n)}\right)
+ \rmi\frac{\erw{p}^{(n)}_\ud z + f^{(n)}_\ud}{\hbar}\right] ,
\end{equation}
with $4 A^{(n)} C^{(n)} - B^{(n)\,2} = \hbar^2$ and
\begin{eqnarray}
f^{(n)}_\ud(t) &= f_0 - \frac{\erw{z}^{(n)}_\ud \erw{p}^{(n)}_\ud}{2} - \frac{\hbar^2}{4m} \int_0^t \frac{\rmd t'}{A^{(n)}(t')} \,.
\end{eqnarray}
With the wave functions \eref{eqn:squeezed-chi-narrow} we can calculate the spin expectation value \eref{eqn:expectation-sigma-x-chi} and find
\numparts
\begin{equation}\label{eqn:spin-expectation-narrow}
 \erw{\hat{\sigma}_x} = \rme^{-\gamma_\mathrm{SN}} 
 \cos \left(\phi_{\alpha\beta} + \varphi_g \right) \,,
\end{equation}
with $\gamma_\mathrm{SN}$ defined as in equation \eref{eqn:gamma-wide-sym}. As before, there is no phase shift due to self-gravity. The only observable effect is in the dampening $\gamma_\mathrm{SN}$.
\endnumparts

For the general case of $\delta \not= 0$, we write again $A^{(1)}_\ud = A^{(0)} + A_+ \pm \delta A_-$, which must satisfy
\begin{equation}\label{eqn:d3Apm}
\partial_t^3 A_\pm = -\frac{\Delta \dot{u}}{m} A^{(0)} k_\pm'(\Delta u) - \frac{\hbar^2 t}{m^3 A_0} k_\pm(\Delta u) \,.
\end{equation}
We find equation \eref{eqn:spin-expectation-asymmetrical} with the phase \eref{eqn:phase-asymmetrical} and the dampening \eref{eqn:gamma-asymmetrical} for the spin expectation value.

\subsection{Self-interaction potential}

\begin{table}
	\begin{center}
		\begin{tabular}{lrrrrr}\hline
			Material &\quad
			$m_a$ / u &\quad
			$\rho$ / \gram \,\power{\centi\meter}{-3} &\quad
			$\sigma$ / \pico\meter &\quad
			$\Wsn$ / \power{\second}{-1} &\quad\quad
			$\wsn$ / \power{\second}{-1} \\
			\hline
			Diamond & 23.011 & 3.520 & 5.71 & 0.044 & $9.93 \times 10^{-4}$ \\
			Silicon & 28.086 & 2.329 & 6.96 & 0.096 & $8.06 \times 10^{-4}$ \\
			Tungsten & 183.84 & 19.30 & 3.48 & 0.695 & $2.31 \times 10^{-3}$ \\
			Osmium & 190.23 & 22.57 & 2.77 & 0.996 & $2.52 \times 10^{-3}$ \\
			Gold & 196.97 & 19.32 & 4.66 & 0.464 & $2.33 \times 10^{-3}$ \\\hline
		\end{tabular}
	\end{center}
	\caption{Atomic mass $m_a$~\cite{oliveReviewParticlePhysics2014}, mass density $\rho$~\cite{oliveReviewParticlePhysics2014}, atomic localization length $\sigma$ at $\unit{0.1}{\kelvin}$~\cite{gaoParameterizationTemperatureDependence1999,searsDebyeWallerFactorElemental1991}, and corresponding frequencies $\Wsn$ and $\wsn$ for selected elements.\label{tab:wsn}}
\end{table}

In order to proceed, we give an explicit form of the self-interaction $I_\rho$ as defined in equation \eref{eqn:irho}.
If the mass distribution $\rho(r)$ is a solid sphere of radius $R$, the integrals can be evaluated and one finds~\cite{iweCoulombPotentialsSpherical1982}
\begin{equation}\label{eqn:I-sphere}
I_{\rho}^\mathrm{sph}(d)=-\frac{G m^2}{R}\times
\left\{ \begin{array}{ll}
\frac{6}{5}-2\left(\frac{d}{2R}\right)^2+\frac{3}{2}\left(\frac{d}{2R}\right)^3-\frac{1}{5}\left(\frac{d}{2R}\right)^5
&\mbox{for}\ d\leq 2R\,,\\
\frac{R}{d}
&\mbox{for}\ d\geq 2R\,.
\end{array}\right.
\end{equation}
However, for a realistic composite particle one must take into account that $\rho$ is peaked around the locations of the constituent atoms and generally has a Gaussian distribution of a width corresponding to the Debye-Waller length $\sigma$~\cite{yangMacroscopicQuantumMechanics2013,grossardtEffectsNewtonianGravitational2016}.
Whereas the mutual gravitational forces between different atoms result in the self-interaction potential $I_{\rho}^\mathrm{sph}$, one must also include the sum of self-interactions for each atom of mass $m_a$ given by
\begin{equation}\label{eqn:I-atom}
 I_\rho^\mathrm{atom}(d) = -\frac{G m m_a}{d}
 \mathrm{erf}\left(\frac{d}{\sqrt{2} \sigma}\right) \,.
\end{equation}
Disregarding the irrelevant constant term in $I_\rho^\mathrm{sph}$, both potentials become comparable for $d \sim \sqrt{m_a/(\rho \sigma)}$, with $I_\rho^\mathrm{atom}$ dominating for smaller distances and becoming negligible as $d$ exceeds said value. For the experimental situations we have in mind, we typically find $R \gg \sqrt{m_a/(\rho \sigma)} \gg \sigma$ of the order of micro-, nano-, and picometers, respectively.

In order to obtain equations of motion that are simple enough to be integrated analytically, yet a good approximation, we thus consider the three intervals from 0 to $\sigma$, to $2R$, and to infinity,
and approximate the derivatives of $I_\rho = I_{\rho}^\mathrm{sph} + I_\rho^\mathrm{atom}$ for each of them separately. We define $\wsn^2 = G m / R^3$ and $\Wsn^2 = \sqrt{2/\pi} G m_a / (3 \sigma^3)$, with values given in table~\ref{tab:wsn} for some materials, and for each of the contributions proportional to $\Wsn$ and $\wsn$, respectively, we only keep the leading order term and we can neglect $I_\rho^\mathrm{atom}$ for $d>2R$. We then find:
\numparts\label{eqnsys:ik}
\begin{eqnarray}
I_\rho'(d) &\approx 
\left\{ \begin{array}{ll}
m \Wsn^2 d + m \wsn^2 d
&\mbox{for}\ d \in [0,\sigma) \\
m \Wsn^2 \sqrt{\frac{\pi}{2}}\frac{3  \sigma^3}{d^2}
+ m \wsn^2 d 
&\mbox{for}\ d \in [\sigma,2R) \\
m \wsn^2 \frac{R^3}{d^2} 
&\mbox{for}\ d \in [2R,\infty) 
\end{array}\right. 
\end{eqnarray}
and for $k_-$ and its derivative 
\begin{eqnarray}
k_-(d) &\approx 
\left\{ \begin{array}{ll}
m \Wsn^2 \frac{9 d^2}{10 \sigma^2}
+ m \wsn^2 \frac{9 d}{8 R}
&\mbox{for}\ d \in [0,\sigma) \\
m \Wsn^2 + m \wsn^2 \frac{9 d}{8 R}
&\mbox{for}\ d \in [\sigma,2R) \\
m \Wsn^2 + m \wsn^2 
&\mbox{for}\ d \in [2R,\infty)  \,.
\end{array}\right. \label{eqn:approx-km} \\
k_-'(d) &\approx 
\left\{ \begin{array}{ll}
m \Wsn^2 \frac{9 d}{5 \sigma^2} + m \wsn^2 \frac{9}{8 R}
&\mbox{for}\ d \in [0,\sigma) \\
-m \Wsn^2 \frac{9 \sqrt{2 \pi} \sigma^3}{d^4}  + m \wsn^2 \frac{9}{8 R}
&\mbox{for}\ d \in [\sigma,2R) \\
-m \Wsn^2 \frac{9 \sqrt{2 \pi} \sigma^3}{d^4}  - m \wsn^2 \frac{6 R^3}{d^4} 
&\mbox{for}\ d \in [2R,\infty)  \label{eqn:approx-kmp} \,.
\end{array}\right.
\end{eqnarray}
\endnumparts
With these approximations, we can then solve
\numparts\label{eqnsys:piecewise-diffeq}
\begin{eqnarray}
\partial_t^2 \Delta z &= -\frac{1}{m} I_\rho'(\Delta u) \label{eqn:piecewise-diffeq-z} \,, \\
\partial_t^3 A_- &= -\frac{\Delta \dot{u}}{m} A^{(0)} k_-'(\Delta u) - \frac{\hbar^2 t}{m^3 A_0} k_-(\Delta u) \label{eqn:piecewise-diffeq-a} \,,
\end{eqnarray}
\endnumparts
piecewise in the different intervals.

\subsection{Spin expectation value}
With the sequence pictured in \fref{fig:setup}, i.\,e. with the solution to \eref{eqn:uddot}, we find that $\Delta u$ reaches a maximum value $\Delta u_\mathrm{max} = a \taua^2 / 2$ after the acceleration time $\taua$. As in the previous section, we can write the resulting spin expectation value after the final time $T = \tau + 2 \taua$ as
\begin{equation}
 \erw{\hat{\sigma}_x} = \sqrt{1-\delta^2} \rme^{-\gamma_\mathrm{SN}^0 - \gamma_\mathrm{SN}^\delta} 
 \cos \left(\phi_{\alpha\beta} + \varphi_g + \Delta\phi_\mathrm{SN} \right) \,,
\end{equation}
where the phase and dampening are obtained as in equations \eref{eqn:phase-asymmetrical} and \eref{eqn:gamma-asymmetrical}.

The solutions of equations \cref{eqnsys:piecewise-diffeq} depend on whether or not $\Delta u_\mathrm{max}$ exceeds the atomic localization scale $\sigma$ and the particle radius $R$, respectively. As a measure for the spreading of the wave function, we introduce the abbreviation
\begin{equation}
 \xi = \frac{\hbar \tau}{2 m A_0} \,.
\end{equation}
In the limiting case where $\taua \ll \tau$, one can then approximate
\numparts\label{eqnsys:localized-gamma0}
\begin{equation}\label{eqn:gamma0}
  \gamma_\mathrm{SN}^0 \approx \frac{\omega_\mathrm{eff}^4 \tau^4 \Delta u_\mathrm{max}^2}{16 \xi A_0}
  \left(\frac{2}{\xi} + \frac{\xi}{2}\right) \,,
\end{equation}
with the effective frequencies
\begin{equation}
\omega_\mathrm{eff}^2 = \left\{ \begin{array}{ll}
\Wsn^2 + \wsn^2
&\quad \mbox{for}\ \Delta u_\mathrm{max} \in [0,\sigma) \\
3 \sqrt{\frac{\pi}{2}} \frac{\sigma^3}{\Delta u_\mathrm{max}^3} \Wsn^2 + \wsn^2 
&\quad \mbox{for}\ \Delta u_\mathrm{max} \in [\sigma,2R) \\
\frac{R^3}{\Delta u_\mathrm{max}^3} \wsn^2
&\quad \mbox{for}\ \Delta u_\mathrm{max} \in [2R,\infty)  \,.
\end{array}\right.
\end{equation}
\endnumparts
The $\delta$ dependent phase and additional dampening are
\numparts
\begin{eqnarray}
\fl \Delta\phi_\mathrm{SN} &\approx \frac{\delta \tau^2}{4} \left(\left(\frac{\xi}{3} - \frac{1}{\xi}\right) \omega_1^2
+ \left(\xi + \frac{1}{\xi} 
- \left(2 + \xi^2\right) \mathrm{arctan}\,\xi \right) \omega_2^2 \right)
+ \delta \, \Delta\phi_\mathrm{int}\\
\fl \gamma_\mathrm{SN}^\delta &\approx \frac{\delta^2 \tau^4}{1 + \xi^2} \left(
\left(\left(\frac{\xi}{6} + \frac{1}{4\xi}\right) \omega_1^2 - \frac{1}{4 \xi} \omega_2^2\right)^2
+ \left(\frac{\xi^2}{12} \omega_1^2\right)^2
\right) \,,
\end{eqnarray}
with the frequencies
\begin{eqnarray}
\fl \omega_1^2 &= \left\{ \begin{array}{ll}
\frac{9 \Delta u_\mathrm{max}^2}{10 \sigma^2} \Wsn^2
+ \frac{9 \Delta u_\mathrm{max}}{8 R} \wsn^2
&\quad \mbox{for}\ \Delta u_\mathrm{max} \in [0,\sigma) \\
\Wsn^2 + \frac{9 \Delta u_\mathrm{max}}{8 R} \wsn^2
&\quad \mbox{for}\ \Delta u_\mathrm{max} \in [\sigma,2R) \\
\Wsn^2 + \wsn^2
&\quad \mbox{for}\ \Delta u_\mathrm{max} \in [2R,\infty)  \,,
\end{array}\right. \\
\fl \omega_2^2 &= \left\{ \begin{array}{ll}
0
&\quad \mbox{for}\ \Delta u_\mathrm{max} \in [0,\sigma) \\
\left(\frac{1}{10} + 6 \sqrt{\frac{\pi}{2}} \left(1 - \frac{\sigma^3}{u_\mathrm{max}^3}\right)\right) \Wsn^2 
&\quad \mbox{for}\ \Delta u_\mathrm{max} \in [\sigma,2R) \\
\left(\frac{1}{10} + 6 \sqrt{\frac{\pi}{2}} \left(1 - \frac{\sigma^3}{u_\mathrm{max}^3}\right)\right) \Wsn^2 
- \left(1 + \frac{2 R^3}{\Delta u_\mathrm{max}^3}\right) \wsn^2
&\quad \mbox{for}\ \Delta u_\mathrm{max} \in [2R,\infty)  \,.
\end{array}\right. 
\end{eqnarray}
\endnumparts
In addition, we need to evaluate the integral phases $\Delta\phi_\mathrm{int}$, which are discussed in \ref{app:fint}. As before, we assume $\taua \ll \tau$. For the narrow separation with $\Delta u_\mathrm{max} \leq \sigma$, they take the simple form $\Delta\phi_\mathrm{int} = - \tau^2 \xi \omega_1^2/6$, resulting in the total phase 
\begin{equation}
 \Delta\phi_\mathrm{SN}^\mathrm{narrow} \approx -\frac{3 \delta \tau^2}{8} \left(\xi
+ \frac{3}{\xi} \right) \left(\frac{\Delta u_\mathrm{max}^2}{5 \sigma^2} \Wsn^2
+ \frac{\Delta u_\mathrm{max}}{4 R} \wsn^2\right)\,.
\end{equation}
For $\Delta u_\mathrm{max} > \sigma$ the integral phases can be expressed by the general form:
\numparts\label{eqnsys:int2phase}
\begin{equation}
\Delta\phi_\mathrm{int} = \frac{\tau^2}{4} \Bigg[
\left(\left(\frac{1}{10} + 6 \sqrt{\frac{\pi}{2}} \kappa_{\Omega} \right) \Wsn^2 + \kappa_{\omega} \wsn^2\right) \zeta(\xi) 
- \frac{2 \xi}{3} \omega_1^2 \Bigg]  \,,
\end{equation}
with
\begin{equation}
\zeta(\xi) = \left(\xi + \frac{1}{\xi}\right) \left(\left(\xi + \frac{1}{\xi}\right) \mathrm{arctan} \, \xi - 1\right)
\end{equation}
and the $\Delta u_\mathrm{max}$ dependent coefficients
\begin{eqnarray}
\kappa_\Omega &\approx \left\{ \begin{array}{ll}
\frac{\sigma^3}{\Delta u_\mathrm{max}^3} - 1
&\quad \mbox{for}\ \Delta u_\mathrm{max} \in [\sigma,2\sigma) \\
1-15 \frac{\sigma^3}{\Delta u_\mathrm{max}^3}
&\quad \mbox{for}\ \Delta u_\mathrm{max} \in [2\sigma,\infty) \,,
\end{array}\right. \\
\kappa_\omega &\approx \left\{ \begin{array}{ll}
\frac{9 (\Delta u_\mathrm{max}-\sigma)}{4 R}
&\quad \mbox{for}\ \Delta u_\mathrm{max} \in [\sigma,2\sigma) \\
\frac{9 \Delta u_\mathrm{max}}{8 R}
&\quad \mbox{for}\ \Delta u_\mathrm{max} \in [2\sigma,2R) \\
3 - \frac{9 \Delta u_\mathrm{max}}{8 R} + \frac{2 R^3}{\Delta u_\mathrm{max}^3} 
&\quad \mbox{for}\ \Delta u_\mathrm{max} \in [2R,4R) \\
-\frac{30 R^3}{\Delta u_\mathrm{max}^3} - 1
&\quad \mbox{for}\ \Delta u_\mathrm{max} \in [4R,\infty)  \,.
\end{array}\right. \label{eqn:localized-phi-int-last}
\end{eqnarray}
\endnumparts
Note that the integral phase changes its sign from negative to positive for some $u_\mathrm{max} \in [\sigma,2\sigma)$, depending on the argument of $\zeta$.

In the above equations \eref{eqnsys:localized-gamma0}--\cref{eqnsys:int2phase}, we have considered the lowest order contributions in the limit $\taua \ll \tau$, while keeping $\Delta u_\mathrm{max}$ constant, implying that the acceleration $a$ scales accordingly with $\taua$.
Due to the approximations made in equations \cref{eqnsys:ik}, the expressions for the different regimes do not transition into each other continuously at the boundaries $\Delta u_\mathrm{max} = \sigma$ and $\Delta u_\mathrm{max} = 2R$, but provide a good approximation if $\Delta u_\mathrm{max}$ is sufficiently far from those values.

Solutions for the phase and dampening can be obtained, in principle, also for the general case without the approximation $\taua \ll \tau$. The analytic expressions will, however, become rather lengthy, and for the estimation of the order of magnitude of observable effects desired here, the approximate equations provided are sufficient. If the exact expressions \eref{eqn:I-sphere} and \eref{eqn:I-atom} are to be used without the approximations in equations \cref{eqnsys:ik}, the differential equations for $A_-$ and $\Delta z$ likely have to be solved numerically. Nonetheless, the methods outlined here are applicable in order to arrive at the results in a specific situation to desired precision.

Finally, the resulting phase and dampening in the regime of an intermediate wave function width, $\sigma < \sqrt{A_0} < 2R$, can be straightforwardly obtained by simply setting $\Wsn = 0$ in the above equations.

\subsection{Wave packet spreading beyond localization length}\label{sec:ntw}

Before we discuss the experimental consequences of our considerations thus far, let us consider one more special case. A wave function that is initially very well localized and, therefore, has a large momentum uncertainty will spread very fast. If $\xi^2 A_0 > \sigma^2$, it will eventually spread beyond the size where $I_\rho^\mathrm{atom}$ contributes significantly. Assuming $\sqrt{A_0} \ll \sigma$ (otherwise we can simply neglect the contributions from $I_\rho^\mathrm{atom}$ altogether), we find the time $\tau_s$ where the wave function reaches this limit to be
\begin{equation}
 \tau_s \approx \frac{2 m}{\hbar} \sqrt{A_0} \sigma
 = \frac{\tau \, \sigma}{\xi \, \sqrt{A_0}} \,.
\end{equation}
We then have to evaluate equations \cref{eqnsys:piecewise-diffeq} with $\Wsn \not= 0$ only for $t < \tau_s$ and $\Wsn = 0$ for later times. We still assume that the wave function is narrow compared to the particle size at all times---otherwise the approximations in this section are no longer valid and we need to also account for the considerations in \sref{sec:gauss}.

In the case $\Delta u_\mathrm{max} > 4 R$, one finds effectively the result for an intermediate wave function where $\Wsn = 0$ with some additional terms, in total:
\numparts
\begin{eqnarray}
\fl \gamma_\mathrm{SN}^0 &\approx \frac{\tau^4 \Delta u_\mathrm{max}^2}{16 \xi A_0}
\left[\left(\frac{2}{\xi} + \frac{\xi}{2}\right) \frac{R^6}{\Delta u_\mathrm{max}^6} \wsn^4
+ \frac{8 \tau^4 \sigma^6}{9 \xi^7 \taua^4 A_0^3} \Wsn^4 \right]\\
\fl \Delta\phi_\mathrm{SN} &\approx -\frac{\delta \tau^2}{4} \Bigg[\left(\frac{\xi}{3} + \frac{1}{\xi}\right) \wsn^2
+ \zeta(\xi) \left(
\sqrt{\frac{\pi}{2}} \frac{42 \sigma^3}{\Delta u_\mathrm{max}^3} \Wsn^2
+ \frac{28 R^3}{\Delta u_\mathrm{max}^3} \wsn^2 \right)
\nnl\fl&\bleq
+\frac{1}{\xi^2} \mathrm{arctan}(\xi)
\left(\frac{18 \tau^4 \sigma^2 \Delta u_\mathrm{max}^2}{5 \xi^4 \taua^4 A_0^2}  \Wsn^2 +
\left(1 + \frac{2 R^3}{\Delta u_\mathrm{max}^3}\right) \wsn^2 \right)
\Bigg] \\
\fl \gamma_\mathrm{SN}^\delta &\approx \frac{\delta^2 \tau^4}{1 + \xi^2} \Bigg[
\left(\left(\frac{\xi}{6} + \frac{1}{2 \xi}\left(1+ \frac{R^3}{\Delta u_\mathrm{max}^3}\right)\right) \wsn^2 
+ \frac{9 \tau^4 \sigma^2 \Delta u_\mathrm{max}^2}{10 \xi^5 \taua^4 A_0^2} \Wsn^2\right)^2
\nnl\fl&\bleq
+ \left(\frac{\xi^2}{12} \wsn^2\right)^2
\Bigg] \,.
\end{eqnarray}
\endnumparts
In the limit $\xi \to \infty$ we can approximate further:
\numparts\label{eqnsys:ntw-approx}
\begin{eqnarray}
\gamma_\mathrm{SN}^0 &\approx \frac{ R^6}{32 A_0 \Delta u_\mathrm{max}^4} \wsn^4\tau^4
\ll \gamma_\mathrm{SN}^\delta \\
\Delta\phi_\mathrm{SN} &\approx -\delta^2 \frac{7  \pi    R^3}{2 \Delta u_\mathrm{max}^3} \xi^2  \wsn^2\tau^2 \label{eqn:ntw-approx-phase} \\
\gamma_\mathrm{SN}^\delta &\approx \delta \frac{\xi^2}{144} \wsn^4\tau^4  \,.
\end{eqnarray}
\endnumparts
One finds that in this limit the short time for which the wave function is narrow compared to the atomic localisation length scale $\sigma$ plays no significant role and one can entirely ignore the contributions from $I_\rho^\mathrm{atom}$. Equations \cref{eqnsys:ntw-approx} are identical to what is obtained from equations \eref{eqnsys:localized-gamma0}--\cref{eqnsys:int2phase} in the limit of large $\xi$ and for $\Delta u_\mathrm{max} > 4R$.

\section{Discussion}\label{sec:discussion}

In a proper treatment of the self-gravitational interaction due to semi-classical gravity, we found expressions for both the phase shift and the dampening of the visibility for interferometric spin measurements. We studied both the case of a point-like particle in a Gaussian state and localized systems. Let us now turn to the discussion of possible experimental tests of these effects.

\subsection{Decoherence due to acceleration noise}

In the previous sections, we treated the external acceleration $\vec g$ as a constant. We found that the phase and loss of visibility from self-gravity decouple from the external phase due to $\vec g$. However, in most situations where the self-gravitational effects are relevant, the external phase will be large. Even small variations of $\vec g$ will then lead to significant changes of the observed spin expectation value, and in repeated measurements the phase shifts will cancel and result in an additional loss of visibility~\cite{grossardtAccelerationNoiseConstraints2020}. 

For a quantitative description, assume that the acceleration in $z$-direction is normal distributed around $g = \erw{g}$ with deviation $\Delta g$ over the time scale $T$ of the experiment. We assume for simplicity (without great impact on the result) that $g$ is a constant during a single experimental run and only varies between repetitions. We must then average the gravitational phase $\sim \cos \nu g$ and find
\begin{eqnarray}
 \erw{\cos \nu g} &= \left(2\pi \Delta g^2\right)^{-1/2} \int \rmd g \, \exp\left(-\frac{(g - \erw{g})^2}{2\Delta g^2}\right) \,
 \cos \nu g 
 \nnl &= \rme^{- \nu^2 \Delta g^2 / 2} \cos \nu \erw{g} \,.
\end{eqnarray}
Hence, there is an additional dampening $\exp(-\gamma_{\Delta g})$ of the phases $\varphi_g$ and $\phi_\mathrm{SN}$ which in the case of the single sequence with \eref{eqn:phig} takes the form
\begin{equation}\label{eqn:gamma-g}
\gamma_{\Delta g} = \frac{m^2 a^2 \, \Delta g^2}{8 \hbar^2} \taua^4 (\tau + \taua)^2
\approx \frac{1}{2} \left(\frac{m \, \Delta g\, \tau \, \Delta u_\mathrm{max} }{\hbar}\right)^2\,,
\end{equation}
with the last equality assuming $\taua \ll \tau$ and $\Delta u_\mathrm{max} = a \taua^2 /2$ as before.

\subsection{Previous experimental proposals}

The scenario depicted in \fref{fig:setup} is based on the proposal by Bose et al.~\cite{boseSpinEntanglementWitness2017}. In order to observe gravitational spin entanglement between two particles, they propose two adjacent Stern-Gerlach interferometers. For the study of the SN effects, a single interferometer is sufficient. The proposal by Bose et al.\ suggests to use a microdiamond of $m = \unit{10^{-14}}{\kilo\gram}$ ($R \approx \unit{0.9}{\micro\meter}$), as well as times $\taua = \unit{0.5}{\second}$ and $\tau = \unit{2.5}{\second}$. The acceleration results from a field gradient of $\partial_z B \approx \unit{10^6}{\tesla\per\meter}$ which amounts to $a \approx \unit{1}{\milli\meter\per\second\squared}$. The suggestion to initially release the particle from $\sim \unit{1}{\mega\hertz}$ traps implies $\sqrt{A_0} \approx \unit{0.1}{\pico\meter}$. 
Hence, at least initially we are very obviously in the localized regime and have $\Delta u_\mathrm{max}\approx \unit{100}{\micro\meter} \gg R$. We find $\xi \approx 1.3 \times 10^6$.

As $\xi^2 A_0 \gg \sigma^2$, we are clearly in the situation discussed in \sref{sec:ntw}.
One finds $\gamma_\mathrm{SN}^0 \approx 6 \times 10^{-7}$ and, therefore, no suppression of visibility due to self-gravity in the symmetric situation $\delta = 0$. If, however, we allow for an asymmetry $\delta \not= 0$, we find both the additional dampening $\gamma_\mathrm{SN}^\delta \approx 0.5 \delta^2$ as well as the phase $\Delta \phi_\mathrm{SN} \approx -87 \delta$ of the order of unity. Note, however, that for $\gamma_{\Delta g}$ to remain below unity one requires $\Delta g \lesssim \unit{4 \times 10^{-17}}{\meter\per\second\squared}$, imposing very strong limits on the allowed acceleration noise~\cite{grossardtAccelerationNoiseConstraints2020}.

Equations \cref{eqnsys:ntw-approx} suggest that the phase becomes the larger, the smaller $\Delta u_\mathrm{max}$. Limiting $\Delta u_\mathrm{max}$ also results in a smaller effect of acceleration noise. Putting the value at the limit $\Delta u_\mathrm{max} \sim 4 R$, where the approximations \cref{eqnsys:ntw-approx} just remain valid and one requires a field gradient $\partial_z B \sim \unit{31}{\kilo\tesla\per\meter}$, we find a phase of $\Delta \phi_\mathrm{SN} \approx -2 \times 10^6 \delta$ with the symmetric $\gamma_\mathrm{SN}^0 \approx 0.4$ now also of the order of unity. The requirement from acceleration noise is loosened to $\Delta g \lesssim \unit{10^{-15}}{\meter\per\second\squared}$.

We find that with the very parameters suggested by Bose et al.\ in order to barely achieve the required spin entanglement, self-gravity already can amount to huge effects, suggesting that feasibility requirements for a test of self-gravity might be orders of magnitude below those to detect gravitational entanglement. We discuss some scenarios below.

In a recent preprint, Hatifi and Durt~\cite{hatifiRevealingSelfgravitySternGerlach2020} pursue the same idea of a test of the SN equation in a Stern-Gerlach interferometer as described by Bose et al.\footnote{The author only became aware of reference~\cite{hatifiRevealingSelfgravitySternGerlach2020} when this present work had already been far advanced; as the employed methods vary between both works and the unpublished work by Hatifi and Durt seems to have substantial shortcomings, we refrain from referencing it elsewhere in the text.} They consider only the situation where the wave function is small compared to the particle radius and correctly assert that an asymmetric superposition is required in order to observe a phase shift. However, their derivation seems fundamentally flawed as the alleged effect is based on the constant contribution of the gravitational self-energy to the potential~\eref{eqn:u-quadratic-origin}. As is evident from our derivations, the constant contribution to $I_\rho$ appears nowhere, only its derivatives enter into any observable quantity. The intuition behind the considerations in reference~\cite{hatifiRevealingSelfgravitySternGerlach2020} seems to be a mental split of the wave function into two particles of masses $\abs{\alpha}^2 m$ and $\abs{\beta}^2 m$, respectively. Then one would obtain differing gravitational self-energies which result in differing overall phases. If the constant self-energy is instead properly included in the derivation, it enters as a constant in the Schrödinger equation~\eref{eqn:se-n}. $\psi_\uparrow$ and $\psi_\downarrow$ then attain equal overall phases which lead to no observable effect.

The experimental parameters suggested by Hatifi and Durt~\cite{hatifiRevealingSelfgravitySternGerlach2020} are a micron sized particle with a mass of $\unit{5.5 \times 10^{-15}}{\kilo\gram}$ with an initial wave function spread of $\sqrt{A_0} = \unit{1}{\nano\meter} \ll R \approx \unit{1}{\micro\meter}$, $\taua = \unit{0.5}{\second}$ as before, and $\tau = \unit{1}{\second}$. The field gradient is supposed to be the same as chosen by Bose et al., which due to the scaling with the inverse mass results in $\Delta u_\mathrm{max} \sim \unit{180}{\micro\meter}$. Despite being small compared to the particle size, the wave function width is actually wider than the atomic localization length scale and we are in the regime where $I_\rho^\mathrm{atom}$ does not contribute and we effectively have $\Wsn = 0$. Hatifi and Durt do not explicitly suggest a material for the microsphere, hence we assume $\wsn \approx \unit{10^{-3}}{\power{\second}{-1}}$. For these values, we find that---besides negligible loss of visibility from self-gravity---the phase of $\Delta \phi_\mathrm{SN} \approx -5 \times 10^{-5}$ is likely unobservable, and deviates significantly from the observable phase claimed in their preprint. The constraints from acceleration noise, on the other hand, still require $\Delta g \lesssim \unit{10^{-16}}{\meter\per\second\squared}$ for an observable phase.

\subsection{Requirements for an observable phase shift}
From the comparison of the two proposals above we see that the particle mass, flight time, and acceleration are not the only relevant parameters. The considerably more localized wave function in the proposal by Bose et al.\ results in a much larger phase.

Intuitively, this is easily understood if one considers the different contributions to the phase by themselves and notices that the dominant contribution, by far, results from the integral in \eref{eqn:phase-f}.
In the case of an asymmetrical superposition, the gravitational potential is split asymmetrically, as well, and the spreading of the wave function slows down differently for the spin-up and spin-down parts. The integrand is proportional to $A^{-1}$, implying that a narrower wave function contributes more strongly.

In order to find a feasible scenario to observe the phase shift, therefore, we start from equation \eref{eqn:ntw-approx-phase}, allow for the maximum spread $\xi^2 A_0 \sim R^2$ and the minimum $\Delta u_\mathrm{max} \sim 4R$ for which this limit does still apply, finding
\begin{equation}
\Delta \phi_\mathrm{SN} \sim - \delta \frac{7 \pi^3 \, \rho^2 \, R^{10} \, \wsn^2}{18 \, \hbar^2} \,.
\end{equation}
With the values in \tref{tab:wsn}, one finds an observable phase for a particle of $R \sim \unit{110}{\nano\meter}$ for silicon or $R \sim \unit{56}{\nano\meter}$ for osmium.
If $\Delta g$ is the largest possible value such that decoherence from acceleration noise does not lead to complete loss of visibility, and we consider $\taua \ll \tau$, we find that the acceleration must be
\begin{equation}
 a \gg \frac{96 m^3 \, \Delta g^2}{\pi \hbar^2 \, \rho} \,.
\end{equation}
This implies that observation is feasible, for instance, for a field gradient $\partial_z B \sim \unit{10^5}{\tesla\per\meter}$ with a flight time $\tau \sim \unit{2}{\milli\second}$ and a particle released from a gigahertz trap, if acceleration noise in the kilohertz range of the spectrum can be kept below $\unit{10^{-8}}{\meter\per\second\squared}$, as achieved in drop tower experiments~\cite{seligDropTowerMicrogravity2010}.

Finally, consider the scenario of a wide wave function with the approximation \eref{eqn:approx-wide-phase} for the phase. The conditions for this approximation amount to $\Delta u_\mathrm{max} \gg \sqrt{A_0} \gg R$. The requirement of negligible spreading implies a maximum time $\tau \sim \rho R^5 / \hbar$, with which one finds the ideal parameters $R \sim \unit{24}{\nano\meter}$ and $\tau \sim \unit{150}{\milli\second}$. With these values, one achieves a phase of the order of unity, which is visible for acceleration noise below $\Delta g \sim \unit{10^{-6}}{\meter\per\second\squared}$.

\subsection{Requirements for an observable loss of visibility}
As the phase shift requires an asymmetrical superposition which may be somewhat more difficult to prepare and measure, it is worth looking into the $\delta$-independent part $\gamma_\mathrm{SN}^0$ of the loss of visibility. If one again requires $\Delta g$ to take the largest possible value to avoid decoherence, equation \eref{eqn:gamma0} can be written
\begin{equation}
\gamma_\mathrm{SN}^0 \approx \frac{\omega_\mathrm{eff}^4 \tau^2 \hbar^2}{32 m^2 \, A_0 \, \Delta g^2}
  \left(1 + \frac{4}{\xi^2}\right) \lesssim \left(\frac{\omega_\mathrm{eff}^2 \, r_\mathrm{eff}}{\Delta g} \right)^2\,,
\end{equation}
where the latter inequality stems from the requirement that the wave function does not spread beyond $r_\mathrm{eff}$.

In the narrow regime, where $I_\rho^\mathrm{atom}$ contributes significantly, we have $\omega_\mathrm{eff} = \Wsn$ and $r_\mathrm{eff} = \sigma$, implying that for an observable loss of visibility, $\gamma_\mathrm{SN}^0 \sim 1$, we must have acceleration noise below $\unit{6\times 10^{-14}}{\meter\per\second\squared}$ for silicon or $\unit{3\times 10^{-12}}{\meter\per\second\squared}$ for osmium.

For wider wave functions up to the particle radius, we require a minimum radius $R \sim \Delta g / \wsn^2$ for any given acceleration noise. However, we must also require $\Delta u_\mathrm{max} \lesssim R$ in order to find the effective frequency not be suppressed, which together with the requirement $\taua \ll \tau$ implies
\begin{equation}
 \Delta g \ll \left(\frac{\mu_B \, \partial_z B \, \tau^2}{\rho}\right)^{1/4} \wsn^2 \,.
\end{equation}
For silicon with $\tau \sim \unit{1}{\second}$ and $\partial_z B \sim \unit{10^6}{\tesla\per\meter}$ this amounts to an acceleration noise below $\unit{5\times 10^{-12}}{\meter\per\second\squared}$.

Finally, let us briefly look at the wide wave function regime with the dampening approximated by equation \eref{eqn:gamma0-wide-approx}. The requirement $\gamma^0_\mathrm{SN} \sim 1$ can be written as
\begin{equation}
 \tau R^5 \sim \frac{9 \hbar}{4 \sqrt{2} \pi^2 G \rho^2} \sim \unit{5 \times 10^{-32}}{\second\,\power{\meter}{5}} \,,
\end{equation}
where acceleration noise needs to remain below $\Delta g \sim \sqrt{2} \pi \, G \rho \, R / 3$. This can be achieved for $\tau \sim \unit{1}{\second}$, $R \sim \unit{550}{\nano\meter}$, and $\Delta g < \unit{10^{-13}}{\meter\per\second\squared}$. The limit on acceleration noise drops proportionally with growing particle radius, resulting to a decreased time $\tau$. However, the requirement $A_0 \gtrsim R^2$ implies a trap frequency $\sim G \rho \tau$, which becomes feasible only for times $\tau \ll \unit{1}{\second}$ and, therefore, poses much stricter constraints on acceleration noise.

\section{Conclusion}\label{sec:conclusion}
Our analysis suggests that a direct test for semi-classical gravitational self-interaction on spin interference poses considerably less constraining requirements on experiments than suggested tests of gravitational entanglement~\cite{boseSpinEntanglementWitness2017,marlettoGravitationallyInducedEntanglement2017}. The requirements may also be easier to achieve than comparable optomechanical tests of self-gravity~\cite{yangMacroscopicQuantumMechanics2013,grossardtOptomechanicalTestSchrodingerNewton2016}.
Interestingly, and contrary to these previous proposals, due to the important role of acceleration noise as a limiting decoherence effect, the best parameter regime appears to be that of wide wave functions for particles of relatively small masses.

Accounting for the fact that alternative suggestions for semi-classical models are either in conflict with observation~\cite{kafriClassicalChannelModel2014,altamiranoGravityNotPairwise2018} or can be tested by other means~\cite{tilloySourcingSemiclassicalGravity2016,donadiUndergroundTestGravityrelated2021}, the direct test proposed here can be considered equivalent to tests for gravitational entanglement, except for the possibility of yet unknown non-entangling, non-self-interacting models.

In order to proceed towards an experimental test of the self-gravitational effects described in this paper, a more detailed study of the relevant parameter regimes is required, including relevant decoherence effects and measurement uncertainties. The methods introduced here can be employed for carrying out these detailed studies. A more precise numerical analysis is desirable, as the parameter regime with the presumably best conditions for an experimental test is, unfortunately, also the one least accessible to good analytical approximations.

\appendix

\section{Schrödinger equation with homogeneous force}\label{app:theorem}
We recall a theorem that clarifies the well known behavior of the Schrödinger equation in a homogeneous potential, namely that the evolution of the wave function is the same as in the rest frame except for a shift of the full wave function, corresponding to its acceleration, and an additional phase:

$\psi(t,\vec r)$ solves the Schrödinger equation
\begin{equation}\label{eqn:app-theo-se-acc-2}
 \rmi \hbar \dot{\psi}(t,\vec r) = -\frac{\hbar^2}{2m} \nabla^2 \psi(t,\vec r) + \left(V(t,\vec r) - \vec F(t) \cdot \vec r\right) \psi(t,\vec r)
\end{equation}
with a homogeneous force $\vec F(t)$ and an arbitrary potential $V(t,\vec r)$, if and only if
\begin{eqnarray}\label{eqn:app-theo-wf-acc-rest-2}
\psi(t,\vec r) &= \rme^{\rmi \varphi(t,\vec r)} \chi(t,\vec r-\vec u(t)) \\
\varphi(t,\vec r) &= \frac{m}{\hbar} \,\vec r \cdot \dot{\vec u}(t) - \frac{m}{2 \hbar} \int_0^t \rmd t' \dot{\vec u}(t')^2\,,
\end{eqnarray}
where $\vec u(t)$ solves the classical equations of motion for the force $\vec F$,
\begin{equation}
\ddot{\vec u}(t) = \frac{\vec F(t)}{m} 
\qquad\mbox{with}\qquad
\vec u(0)= \dot{\vec u}(0) = 0 \,,
\end{equation}
and $\chi(t,\vec r)$ solves the Schrödinger equation in the rest frame with shifted potential,
\begin{equation}\label{eqn:app-theo-se-rest-2}
 \rmi \hbar \dot{\chi}(t,\vec r) = -\frac{\hbar^2}{2m} \nabla^2 \chi(t,\vec r) + V(t,\vec r + \vec u(t)) \chi(t,\vec r)\,,
\end{equation}
with the same initial conditions $\chi(0,\vec r) = \psi(0,\vec r)$.

The proof is a straightforward calculation and left as an exercise for the reader's entertainment.

\section{Self-gravitational potential}\label{app:potential}
We calculate the potential \eref{eqn:self-grav-pot-gauss} for a Gaussian wave function:
\begin{eqnarray}
\fl \widetilde{U}(t,\vec r) &= \left(2 \pi A\right)^{-3/2} \int \rmd^3 r' \,
\exp\left(-\frac{ {r'}^2}{2 A}\right) I_\rho(\vec r - \vec r')
\nnl\fl &=
\frac{1}{\sqrt{2 \pi A^3}} \int_0^\infty \rmd r' \, r'^2 \,
\exp\left(-\frac{{r'}^2}{2 A}\right) \int_{-1}^{1} \rmd u \,I_\rho(\sqrt{r^2 + {r'}^2 - 2 r r' u})
\nnl\fl &=
\frac{1}{\sqrt{2 \pi A^3}} \int_0^\infty \rmd r' \, \frac{r'}{r} \,
\exp\left(-\frac{{r'}^2}{2 A}\right) \int_{\abs{r-r'}}^{\abs{r+r'}} \rmd s \,s \,I_\rho(s)
\end{eqnarray}
For a solid sphere with \eref{eqn:I-sphere} one finds
\begin{eqnarray}
\fl\int_{\abs{r-r'}}^{\abs{r+r'}} \rmd s \,s \,I_\rho(s) &=
-2 G m^2 R \int_{\abs{r-r'}/(2R)}^{\abs{r+r'}/(2R)} 
\left\{ \begin{array}{ll}
\frac{12}{5}x-4 x^3+3 x^4-\frac{2}{5} x^6
&\mbox{for}\ x\leq 1\\
1
&\mbox{for}\ x\geq 1
\end{array}\right\} \rmd x
\nnl\fl &=
-2 G m^2 R
\left\{ \begin{array}{ll}
I_1(R,r,r')
&\mbox{for}\ \abs{r+r'}\leq 2R\,,\\
I_2(R,r,r')
&\mbox{for}\ \abs{r-r'}\leq 2R < \abs{r+r'}\,,\\
I_3(R,r,r')
&\mbox{for}\ 2R < \abs{r-r'}\,.
\end{array}\right.
\end{eqnarray}
with
\begin{eqnarray}
I_1 &= \int_{\abs{r-r'}/(2R)}^{\abs{r+r'}/(2R)} \left(\frac{12}{5}x-4 x^3+3 x^4-\frac{2}{5} x^6\right) \rmd x \\
I_2 &= \int_{\abs{r-r'}/(2R)}^{1} \left(\frac{12}{5}x-4 x^3+3 x^4-\frac{2}{5} x^6\right) \rmd x + \int_{1}^{\abs{r+r'}/(2R)} \rmd x \\
I_3 &= \int_{\abs{r-r'}/(2R)}^{\abs{r+r'}/(2R)} \rmd x
\end{eqnarray}
For $r\leq R$ we then must integrate
\begin{eqnarray}
\fl \widetilde{U}(t,\vec r) &= 
-\frac{1}{\sqrt{2 \pi A^3}}\frac{2 G m^2 R}{r}
\Bigg[ \int_0^{2R-r} \rmd r' \, r' \,
\exp\left(-\frac{{r'}^2}{2 A}\right) I_1 \nnl
\fl &\bleq + \int_{2R-r}^{2R+r} \rmd r' \, r' \,
\exp\left(-\frac{{r'}^2}{2 A}\right) I_2
+ \int_{2R+r}^\infty \rmd r' \, r' \,
\exp\left(-\frac{{r'}^2}{2 A}\right) I_3 \Bigg] \,,
\end{eqnarray}
for $R < r \leq 2R$ we integrate
\begin{eqnarray}
\fl \widetilde{U}(t,\vec r) &= 
-\frac{1}{\sqrt{2 \pi A^3}}\frac{2 G m^2 R}{r}
\Bigg[ \int_0^{2R-r} \rmd r' \, r' \,
\exp\left(-\frac{{r'}^2}{2 A}\right) I_1 \nnl
\fl &\bleq + \int_{2R-r}^{2R+r} \rmd r' \, r' \,
\exp\left(-\frac{{r'}^2}{2 A}\right) I_2
+ \int_{2R+r}^\infty \rmd r' \, r' \,
\exp\left(-\frac{{r'}^2}{2 A}\right) I_3\Bigg] \,,
\end{eqnarray}
whereas for $r>2R$ we integrate
\begin{eqnarray}
\fl \widetilde{U}(t,\vec r) &= 
-\frac{1}{\sqrt{2 \pi A^3}}\frac{2 G m^2 R}{r}
\Bigg[ \int_0^{r-2R} \rmd r' \, r' \,
\exp\left(-\frac{{r'}^2}{2 A}\right) I_3 \nnl
\fl &\bleq + \int_{r-2R}^{r+2R} \rmd r' \, r' \,
\exp\left(-\frac{{r'}^2}{2 A}\right) I_2
+ \int_{r+2R}^\infty \rmd r' \, r' \,
\exp\left(-\frac{{r'}^2}{2 A}\right) I_3\Bigg] \,,
\end{eqnarray}
which are all equal.
In the limit $R \ll \sqrt{A}$ we find
\begin{eqnarray}\label{eqn:app-small-r-approx-for-u}
 \widetilde{U}(t,r) &\approx -\frac{G m^2}{r} \left( 
 \mathrm{erf}\left(\sqrt{\frac{r^2}{2 A}}\right)
 -\frac{1}{5} \sqrt{\frac{2}{\pi A^3}} r R^2 \rme^{-\frac{r^2}{2A}}
 \right) + \order{R^4/A^2} \,.
\end{eqnarray}

\section{Expectation values of gravitational potential}\label{app:expvals}
For the potential
\begin{equation}\label{eqn:app-small-r-approx-for-u-2}
\widetilde{U}(t,r) = -\frac{G m^2}{r} 
 \mathrm{erf}\left(\sqrt{\frac{r^2}{2 A}}\right)
\end{equation}
we calculate the expectation values \eref{eqn:expval-pz}, \eref{eqn:expval-B}, and \eref{eqn:expval-C} in the zeroth order wave function
\begin{equation}\label{eqn:app-gaussian-wf}
\abs{\psi(t,r)}^2 = \left(2 \pi A\right)^{-3/2}
\exp\left(-\frac{r^2}{2 A}\right)
\end{equation}
Note that we only need to consider the real valued absolute value of the wave function, since we will only consider expectation values of functions of position, not momentum.
The potential
\begin{equation}
V_{\vec s}(t,\vec r) = \widetilde{U}(t,\abs{\vec r + \vec s})
= \int \rmd^3 r' \abs{\psi(t,\vec r')}^2 I_\rho(\vec r + \vec s - \vec r') 
\end{equation}
has the symmetries
\begin{equation}
\erw{V_{-\vec s}} = \erw{V_{\vec s}} \,,\qquad
\erw{\partial_z V_{-\vec s}} = -\erw{\partial_z V_{\vec s}} \,,\qquad
\erw{z \partial_z V_{-\vec s}} = \erw{z \partial_z V_{\vec s}} \,.
\end{equation}
With the probability current
\begin{equation}
 \vec j = \frac{\rmi \hbar}{2 m} \left(\psi \nabla \psi^* - \psi^* \nabla \psi\right) \,,
 \qquad \Leftrightarrow \qquad
 2 \psi^* \nabla \psi = \nabla \abs{\psi}^2 + \frac{2\rmi m}{\hbar} \vec j \,,
\end{equation}
and the continuity equation
\begin{equation}
 \nabla \cdot \vec j = -\partial_t \abs{\psi(t,\vec r)}^2 \,,
\end{equation}
we can show (cf. also the supplemental material of reference \cite{grossardtEffectsNewtonianGravitational2016})
\begin{eqnarray}
\fl \erw{(\nabla V_{\vec s}) \cdot \vec p + \vec p \cdot \nabla V_{\vec s}} &=
-\rmi \hbar \int \rmd^3 r \int \rmd^3 r' \Big[ \abs{\psi(t,\vec r)}^2 \abs{\psi(t,\vec r')}^2 \nabla^2 I(\vec r + \vec s - \vec r')
\nnl \fl &\bleq + 2 \abs{\psi(t,\vec r')}^2 \psi^*(t,\vec r) (\nabla \psi(t,\vec r)) \cdot \nabla I(\vec r + \vec s - \vec r') \Big] \nnl
\fl &= -2m \int \rmd^3 r \int \rmd^3 r' \abs{\psi(t,\vec r')}^2 I(\vec r + \vec s - \vec r') \nabla \cdot \vec j
\nnl
\fl &= 2m \int \rmd^3 r \int \rmd^3 r' \abs{\psi(t,\vec r')}^2 \partial_t \abs{\psi(t,\vec r)}^2 I(\vec r + \vec s - \vec r')
\nnl
\fl &= 2m \erw{\partial_t V_{-\vec s}} 
\nnl
\fl &= m \partial_t \erw{V_{\vec s}}\,,
\end{eqnarray}
where the last step is due to the symmetry of the wave function,
and equivalently, if the wave function is separable and $\vec s$ in $z$-direction,
\begin{equation}
  \erw{(\partial_z V_s) p + p \partial_z V_s} = m \partial_t \erw{V_s} \,.
\end{equation}
We evaluate the time derivative and gradient of the potential \eref{eqn:app-small-r-approx-for-u-2} in the limit $R \to 0$ by first introducing the dimensionless variables
\begin{equation}
\fl \rho^2 = \frac{r^2}{2 A} \,,\qquad
\sigma^2 = \frac{s^2}{2 A} \,,\qquad
\xi^2 = \frac{(\vec r + \vec s)^2}{2 A}
= \rho^2 + \sigma^2 + 2 u \rho \sigma \,,
\end{equation}
with which we find
\begin{eqnarray}
V_s &= -\frac{G m^2}{\sqrt{2 A}} f(\xi) \,,
\qquad f(\xi) =
\frac{\mathrm{erf}\left(\xi\right)}{\xi} \,,\\
\partial_x V_s &= \frac{G m^2}{2 A}
\frac{ \sqrt{1-u^2} \rho \cos \varphi}{\xi} f'(\xi)\,,\\
\partial_z V_s &= \frac{G m^2}{2 A}
\frac{u \rho + \sigma}{\xi} f'(\xi)\,,\\
\vec r \cdot \nabla V_s &= \frac{G m^2}{\sqrt{2 A}}
\frac{\rho^2 + u \rho \sigma}{\xi} f'(\xi)\,,\\
x \partial_x V_s &= \frac{G m^2}{\sqrt{2 A}}
\frac{ (1-u^2) \rho^2 \cos^2 \varphi}{\xi} f'(\xi)\,,\\
z \partial_z V_s &= \frac{G m^2}{\sqrt{2 A}}
\frac{ u^2 \rho^2 + u \rho \sigma}{\xi} f'(\xi)\, \,.
\end{eqnarray}
We first define the functions
\begin{equation}
g_\sigma^\pm(\rho) = \frac{\rme^{-\rho^2}}{\sqrt{\pi}} \left(\mathrm{erf}(\rho + \sigma) \pm \mathrm{erf}(\rho - \sigma)\right)
\end{equation}
with the derivatives
\begin{eqnarray}
{g_\sigma^+}'(\rho) &= \frac{4}{\pi} \rme^{-2 \rho^2 - \sigma^2} \cosh(2 \rho \sigma) - 2 \rho g_\sigma^+(\rho) \\
{g_\sigma^-}'(\rho) &= -\frac{4}{\pi} \rme^{-2 \rho^2 - \sigma^2} \sinh(2 \rho \sigma) - 2 \rho g_\sigma^-(\rho)
\end{eqnarray}
and the integrals
\begin{eqnarray}
\fl I_n^\pm(\sigma) &= \int_0^\infty \rmd \rho \, \rho^n g_\sigma^\pm(\rho) \nnl
\fl &= \left. \frac{\rho^{n+1}}{n+1} g_\sigma^\pm(\rho) \right\vert_0^\infty - \int_0^\infty \rmd \rho \, \frac{\rho^{n+1}}{n+1} {g_\sigma^\pm}'(\rho) \\
\fl I_n^+(\sigma) &= -\frac{4 \rme^{- \sigma^2}}{\pi (n+1)} \int_0^\infty \rmd \rho \, \rho^{n+1} \rme^{-2 \rho^2} \cosh(2 \rho \sigma) + \frac{2}{n+1} I_{n+2}^-(\sigma) \nnl
\fl &= - \frac{\rme^{- \sigma^2}}{\pi \sqrt{2^{n+1}} (1+n)} \Gamma\left(1+\frac{n}{2}\right) M\left(1+\frac{n}{2},\frac{1}{2},\frac{s^2}{2}\right) + \frac{2}{n+1} I_{n+2}^-(\sigma) \\
\fl I_n^-(\sigma) &= \frac{4 \rme^{- \sigma^2}}{\pi (n+1)} \int_0^\infty \rmd \rho \, \rho^{n+1} \rme^{-2 \rho^2} \sinh(2 \rho \sigma) + \frac{2}{n+1} I_{n+2}^-(\sigma) \nnl
\fl &=  \frac{\rme^{- \sigma^2} \sigma}{\pi \sqrt{2^{n+1}}} \Gamma\left(\frac{n+1}{2}\right) M\left(\frac{n+3}{2},\frac{3}{2},\frac{\sigma^2}{2}\right) + \frac{2}{n+1} I_{n+2}^-(\sigma)
\end{eqnarray}
where $M$ is Kummer's hypergeometric function of the first kind and $\Gamma$ the gamma function. This allows to recursively write all integrals in terms of $I_0$ and $I_1$, specifically
\begin{eqnarray}
\fl I_1^+(\sigma) &= \frac{\rme^{-\sigma^2/2}}{\sqrt{2 \pi}} \\
\fl I_1^-(\sigma) &= \frac{\mathrm{erf}(\sigma)}{\sqrt{\pi}} - \frac{\rme^{-\sigma^2/2} \mathrm{erf}(\sigma/\sqrt{2})}{\sqrt{2 \pi}} \\
\fl I_2^-(\sigma) &= \frac{1}{2} I_0^-(\sigma) - \frac{\rme^{-\sigma^2/2}\sigma}{2 \sqrt{2 \pi}} \\
\fl I_3^-(\sigma) &= I_1^-(\sigma) - \frac{\rme^{-\sigma^2}\sigma}{4 \pi}
-\frac{\rme^{-\sigma^2/2} \mathrm{erf}(s/\sqrt{2}) (1+\sigma^2)}{4 \sqrt{2 \pi}} \nnl 
\fl &= \frac{\mathrm{erf}(\sigma)}{\sqrt{\pi}} 
- \frac{\rme^{-\sigma^2}\sigma}{4 \pi}
- \frac{\rme^{-\sigma^2/2} \mathrm{erf}(\sigma/\sqrt{2}) (5+\sigma^2)}{4 \sqrt{2 \pi}}   \\
\fl I_4^-(\sigma) &= \frac{3}{2} I_2^-(\sigma) - \frac{\rme^{-\sigma^2/2}\sigma(3+\sigma^2)}{8 \sqrt{2 \pi}} \nnl 
\fl &= \frac{3}{4} I_0^-(\sigma) - \frac{\rme^{-\sigma^2/2}\sigma(9+\sigma^2)}{8 \sqrt{2 \pi}} \,.
\end{eqnarray}
For the Gaussian wave function \eref{eqn:app-gaussian-wf}
we then find the expectation values
\begin{eqnarray}
\fl \erw{V_s} &= -\frac{G m^2}{\sqrt{2 A}} \frac{2}{\sqrt{\pi}} \int_0^\infty \rmd \rho \,\rho^2\rme^{-\rho^2} \int_{-1}^{1} \rmd u \, f(\xi) \nnl\fl
&= -\sqrt{\frac{2}{\pi A}}  \frac{G m^2}{\sigma} \int_0^\infty \rmd \rho \,\rho\rme^{-\rho^2} \int_{\abs{\rho-\sigma}}^{\rho+\sigma} \rmd \xi \, \mathrm{erf}(\xi) \nnl\fl
&= \sqrt{\frac{2}{A}}  \frac{G m^2}{\sigma} \left[ \int_0^\infty \rmd \rho \frac{2\rho\sinh(2 \rho \sigma)}{\pi \rme^{2\rho^2+\sigma^2}}
- I_2^-(\sigma) - \sigma I_1^+(\sigma) \right]
\nnl\fl
&= -\frac{G m^2}{\sqrt{2A}} \frac{I_0^-(\sigma)}{\sigma}  
\\
\fl \erw{\partial_z V_s} &= \frac{G m^2}{2 A} \frac{2}{\sqrt{\pi}} \int_0^\infty \rmd \rho \,\rho^2\rme^{-\rho^2} \int_{-1}^{1} \rmd u \, \frac{u \rho + \sigma}{\xi} f'(\xi) \nnl\fl
&= \frac{G m^2}{2\sqrt{\pi} A \sigma^2} \int_0^\infty \rmd \rho \,\rho\rme^{-\rho^2} \int_{\abs{\rho-\sigma}}^{\rho+\sigma} \rmd \xi \, (\xi^2 - \rho^2 + \sigma^2) f'(\xi) \nnl\fl
&= \frac{G m^2}{\sqrt{\pi} A \sigma} 
\int_0^\infty \rmd \rho \, \rho \rme^{-\rho^2} \left[\mathrm{erf}(\rho+\sigma) + \mathrm{erf}(\rho-\sigma) - \frac{1}{\sigma} \int_{\abs{\rho-\sigma}}^{\rho+\sigma} \rmd \xi \, \mathrm{erf}(\xi)\right]\nnl
\fl &= \frac{G m^2}{A \sigma^2} 
\left[ \int_0^\infty 
\rmd \rho \, \frac{2\rho\sinh(2 \rho \sigma)}{\pi \rme^{2\rho^2+\sigma^2}}
- I_2^-(\sigma) \right] \nnl \fl
&= \frac{G m^2}{A \sigma} \left( \frac{\rme^{-\sigma^2/2}}{\sqrt{2 \pi}} -\frac{I_0^-(\sigma)}{2 \sigma} \right)
\label{eqn:app-erw-dz}
\\
\fl \erw{\vec r \cdot \nabla V_s} &= \frac{G m^2}{\sqrt{2 A}} \frac{2}{\sqrt{\pi}} \int_0^\infty \rmd \rho \,\rho^2\rme^{-\rho^2} \int_{-1}^{1} \rmd u \, \frac{\rho^2 + u \rho \sigma}{\xi} f'(\xi) \nnl\fl
&= \frac{G m^2}{\sqrt{2 \pi A} \sigma} \int_0^\infty \rmd \rho \,\rho\rme^{-\rho^2} \int_{\abs{\rho-\sigma}}^{\rho+\sigma} \rmd \xi \, (\xi^2 + \rho^2 - \sigma^2) f'(\xi) \nnl\fl
&= \frac{2 G m^2}{\sqrt{2 \pi A} \sigma} \int_0^\infty \rmd \rho \,\frac{\rho^2}{\rme^{\rho^2}}
\left[\mathrm{erf}(\rho+\sigma) - \mathrm{erf}(\rho-\sigma) - \frac{1}{\rho}\int_{\abs{\rho-\sigma}}^{\rho+\sigma} \rmd \xi \, \mathrm{erf}(\xi) \right] \nnl \fl
&= \frac{\sqrt{2} G m^2}{\sqrt{A} \sigma} 
\left[ \int_0^\infty 
\rmd \rho \, \frac{2\rho\sinh(2 \rho \sigma)}{\pi \rme^{2\rho^2+\sigma^2}}
- \sigma I_1^+(\sigma) \right] \nnl \fl
&= -\frac{G m^2 \rme^{-\sigma^2/2}}{2 \sqrt{\pi A}}
\\
\fl \erw{z \partial_z V_s} &= \frac{G m^2}{\sqrt{2 A}} \frac{2}{\sqrt{\pi}} \int_0^\infty \rmd \rho \,\rho^2\rme^{-\rho^2} \int_{-1}^{1} \rmd u \, \frac{u^2 \rho^2 + u \rho \sigma}{\xi} f'(\xi) \nnl\fl
&= \frac{G m^2}{2\sqrt{2 \pi A} \sigma^3} \int_0^\infty \rmd \rho \,\rho\rme^{-\rho^2} \int_{\abs{\rho-\sigma}}^{\rho+\sigma} \rmd \xi \, (\xi^4 - 2 \xi^2 \rho^2 + \rho^4 - \sigma^4) f'(\xi) \nnl\fl
&= \frac{\sqrt{2} G m^2}{\sqrt{\pi A} \sigma} \int_0^\infty \rmd \rho \,\frac{\rho^2}{\rme^{\rho^2}}
\left[\mathrm{erf}(\rho+\sigma) - \mathrm{erf}(\rho-\sigma) - \int_{\abs{\rho-\sigma}}^{\rho+\sigma} \rmd \xi \, \frac{\xi^2 - \rho^2}{\rho \sigma^2} \mathrm{erf}(\xi) \right] \nnl \fl
&= \frac{\sqrt{2} G m^2}{3 \sqrt{A} \sigma^3} \Bigg[ 
\int_0^\infty \rmd \rho \, \frac{2\rho\sinh(2 \rho \sigma)}{\pi \rme^{2\rho^2+\sigma^2}} (1 + \sigma^2 - 2 \rho^2)
- \int_0^\infty \rmd \rho \, \frac{4\rho^2 \sigma \cosh(2 \rho \sigma)}{\pi \rme^{2\rho^2+\sigma^2}} \nnl \fl
&\bleq - \sigma^3 I_1^+(\sigma) + 2 I_4^-(\sigma)\Bigg] \nnl \fl
&= \frac{G m^2}{\sqrt{2 A} \sigma^3} \left(I_0^-(\sigma) - \frac{\rme^{-\sigma^2/2} \sigma (2+\sigma^2)}{\sqrt{2 \pi}} \right)
\label{eqn:app-erw-zdz}
\\
\fl \erw{z \partial_z V_0} &= \frac{G m^2}{\sqrt{2 A}} \frac{2}{\sqrt{\pi}} \int_0^\infty \rmd \rho \,\rho^2\rme^{-\rho^2} \int_{-1}^{1} \rmd u \, u^2 \rho f'(\rho) \nnl\fl
&= \frac{2 \sqrt{2} G m^2}{3 \sqrt{\pi A}} \int_0^\infty \rmd \rho \,\rho^3\rme^{-\rho^2} f'(\rho) \nnl\fl
&= -\frac{G m^2}{6 \sqrt{\pi A}}
\end{eqnarray}

\begin{figure}
 \centering
 \includegraphics[scale=0.53]{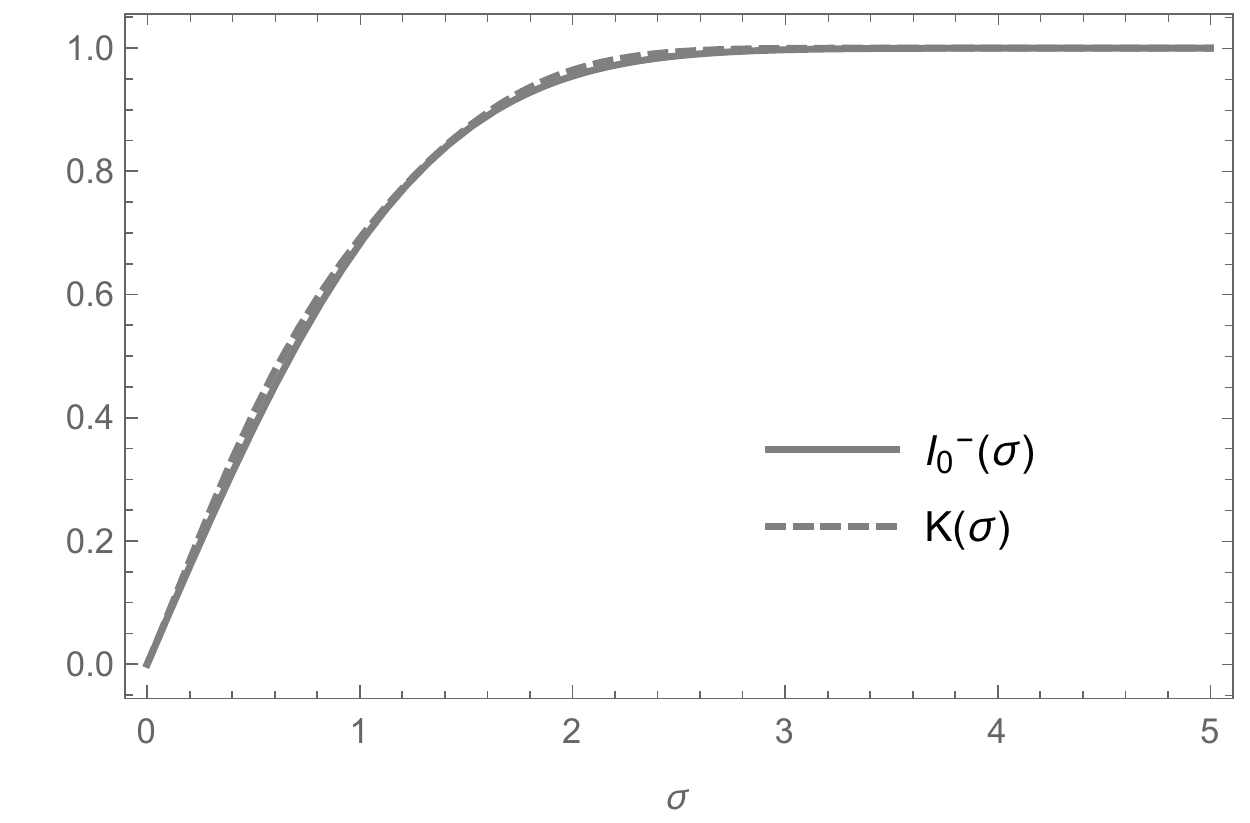}\qquad
 \includegraphics[scale=0.55]{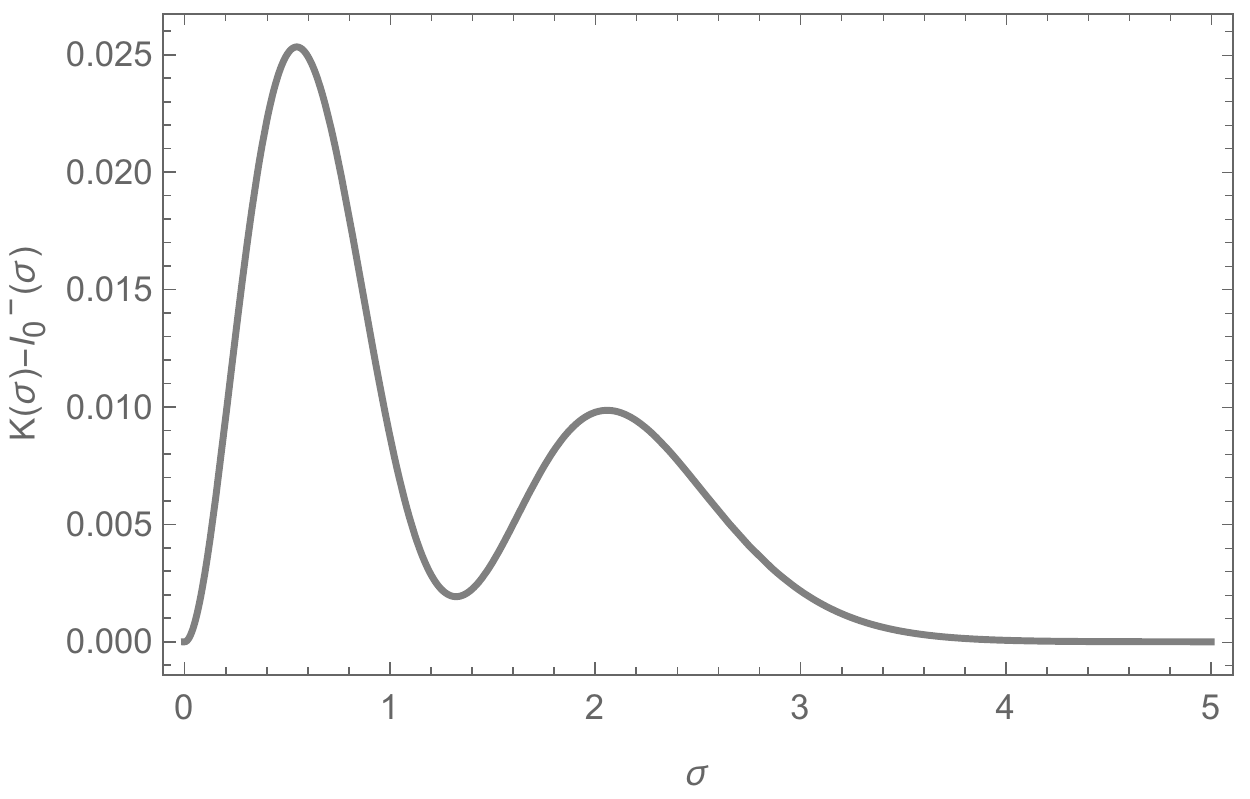}
\caption{Difference between the exact form of $I_0^-(\sigma)$, as per numerical integration, and the approximation $K(\sigma)$. The left pane shows both functions for $\sigma$-values reaching from 0 to 5. The right pane shows the difference between both plots, which tends to zero for large $\sigma$.\label{fig:app-approx-i0}}
\end{figure}
\begin{figure}
 \centering
 \includegraphics[scale=0.55]{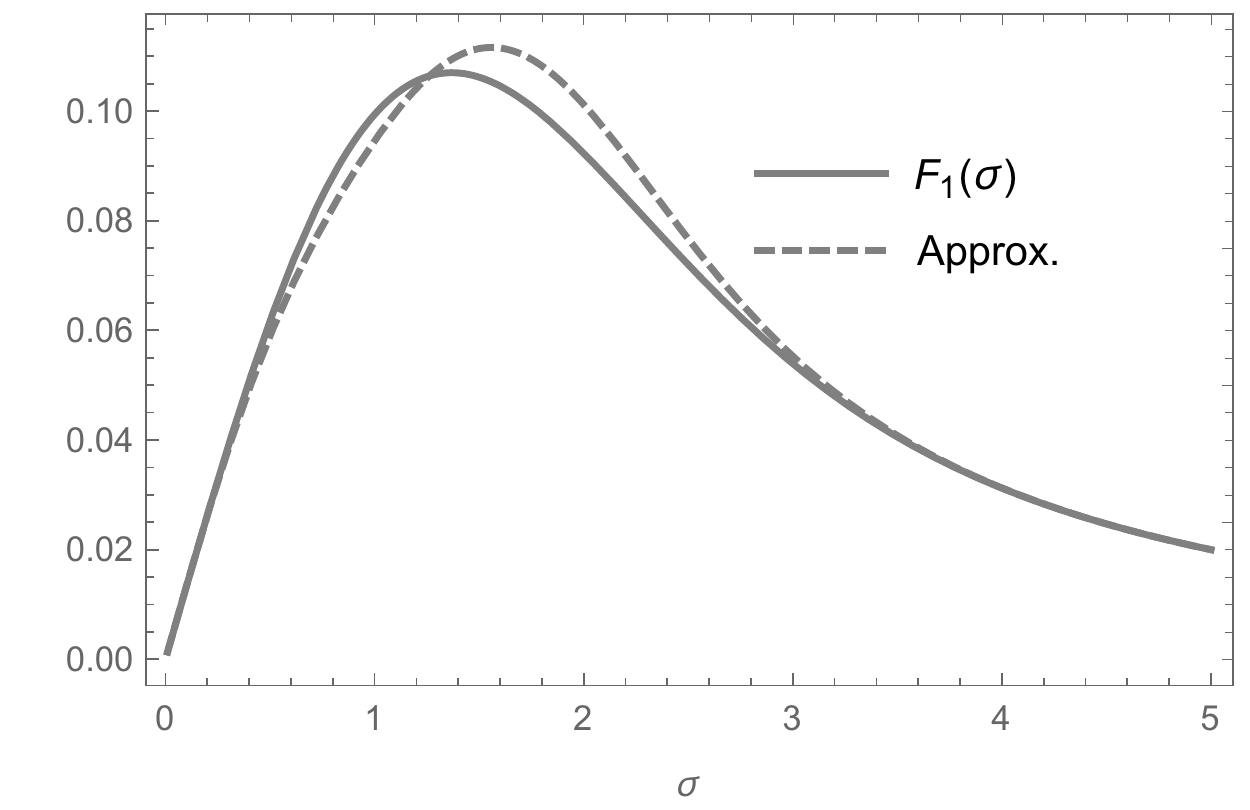}\qquad
 \includegraphics[scale=0.55]{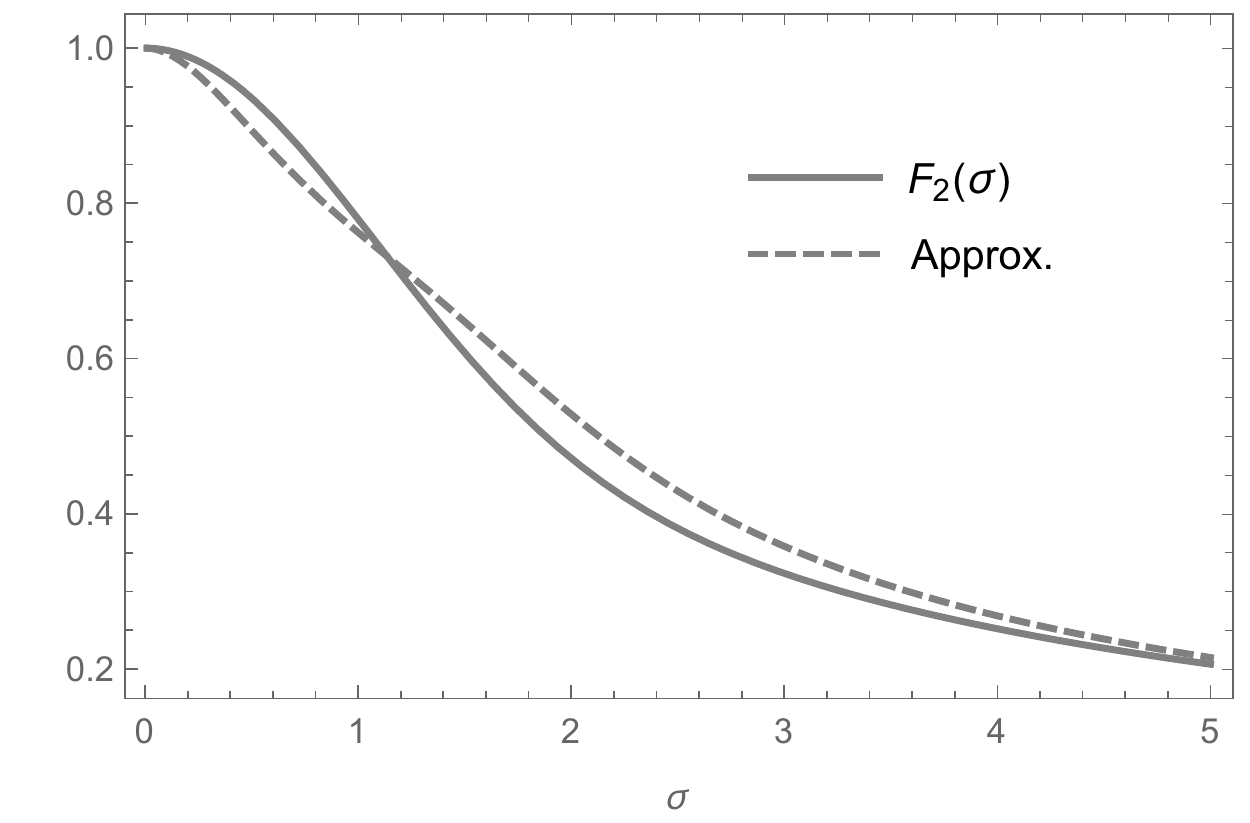}
\caption{Difference between the exact form and the approximation of $F_1$ (left pane) and $F_2$ (right pane).\label{fig:app-approx-f12}}
\end{figure}

We did not succeed at integrating $I_0^-$ analytically. Although it turns out (cf. \fref{fig:app-approx-i0}) that the integral itself is well approximated by the function
\begin{equation}
 K(\sigma) = 1 - \rme^{-\sigma^2} \left(1 - \sqrt{\frac{2}{\pi}} \sigma + \frac{2}{\pi} \sigma^2\right) \,,
\end{equation}
which shows the same behavior in the limits $\sigma \to 0$ and $\sigma \to \infty$, this approximation does not result in the correct limiting behavior for the important functions
\begin{eqnarray}
F_1(\sigma) &= \frac{I_0^-(\sigma)}{2 \sigma^2} - \frac{\rme^{-\sigma^2/2}}{\sqrt{2\pi} \sigma} \\
F_2(\sigma) &= \frac{6}{14 \sigma^3} \left((\sigma^3 + 2 \sigma) \rme^{-\sigma^2/2}
+ (\sigma^2 - 1) \sqrt{2\pi} I_0^-(\sigma)\right) \,.
\end{eqnarray}
found in the main text as determining the evolution equations for the first and second moments. These functions have the limiting behavior
\begin{eqnarray}
F_1(\sigma) &\sim \frac{\sigma}{3 \sqrt{2\pi}} \qquad &\mbox{as } \sigma \to 0 \\
F_1(\sigma) &\sim \frac{1}{2 \sigma^2} \qquad &\mbox{as } \sigma \to \infty \\
F_2(\sigma) &\sim 1 \qquad &\mbox{as } \sigma \to 0 \\
F_2(\sigma) &\sim \frac{3 \sqrt{2\pi}}{7 \sigma} \qquad &\mbox{as } \sigma \to \infty \,.
\end{eqnarray}
We can approximate these functions with the correct limiting behavior by
\begin{eqnarray}
\fl F_1(\sigma) &\approx \frac{1}{2 \sigma^2} \left( 1 - \rme^{-\sigma^2} \left(1 + \sigma^2 - \frac{1}{3}\sqrt{\frac{2}{\pi}} \sigma^3 + \kappa_1 \sigma^4 \right)\right)
 \qquad &\mbox{with } \kappa_1 \approx 0.47
\\
\fl F_2(\sigma) &\approx \frac{3 \sqrt{2\pi}}{7 \sigma} \left( 1 - \rme^{-\sigma^2} \left(1 - \frac{7 \sigma}{3 \sqrt{2 \pi}} + \sigma^2 - \kappa_2 \sigma^3 \right)\right)
 \qquad &\mbox{with } \kappa_2 \approx 0.28 \,,
\end{eqnarray}
where the values for $\kappa_{1,2}$ are obtained by numerical minimization of the integrated squared difference between the approximation and the exact function. The functions and their approximations are plotted in \fref{fig:app-approx-f12}.

\section{Integral contribution to the phase}\label{app:fint}
The phase difference $\Delta \phi_\mathrm{SN}$ acquires a contribution from the integral in \eref{eqn:phase-f}:
\begin{equation}
\Delta f_\mathrm{int} = - \frac{\hbar}{4m} \int_0^T \rmd t \left(\frac{1}{A_\uparrow(t)} - \frac{1}{A_\downarrow(t)}\right) 
\approx \frac{\hbar}{2m} \int_0^T \rmd t \frac{A_-(t)}{A^{(0)}(t)^2}  \,,
\end{equation}
where we made use of the representation \eref{eqn:Aud-to-Apm} and expanded to linear order in $G$. We can write
\begin{equation}
\frac{\hbar}{2 m A^{(0)}(t)^2} = \partial_t^3 Z(t) 
\end{equation}
with
\begin{equation}
Z(t) = \frac{m t}{2 \hbar}\left[ \left(\frac{2 A_0 m}{\hbar t} + \frac{\hbar t}{2 m A_0}\right) 
\mathrm{arctan}\left(\frac{\hbar t}{2 m A_0}\right)
- 1 \right] \,.
\end{equation}
Hence, after threefold integration by parts, we find
\begin{equation}
 \Delta f_\mathrm{int} 
 = \left. A_- \ddot{Z} - \dot{A}_- \dot{Z} + \ddot{A}_- Z \right\vert_0^T + \Delta \phi_\mathrm{int} \,.
\end{equation}
We substitute the time variable by $x(t) = \hbar t / (2 m A_0)$ with $X = x(T)$ and find
\begin{eqnarray}
\Delta \phi_\mathrm{int} 
&= - \int_0^T (\partial_t^3 {A}_-) Z(t) \,\rmd t \nnl
&= \int_0^X k_1(x) \frac{x^3}{1+x^2} \zeta(x) \,\rmd x 
+ \int_0^X k_2(x) x^2 \zeta(x) \,\rmd x \,,
\end{eqnarray}
where we used equation \eref{eqn:d3Apm} with
\begin{eqnarray}
k_1(x) &= \frac{4 m A_0^2}{\hbar^2} k_-(\Delta u) \\
k_2(x) &= \frac{2 m^2 A_0^3}{\hbar^3} \Delta \dot{u} k_-'(\Delta u) \\
\zeta(x) &= \left(x + \frac{1}{x}\right)\left(\left(x + \frac{1}{x}\right) \mathrm{arctan} x - 1\right) \,.
\end{eqnarray}
Solving equation \eref{eqn:uddot}, defining $\xi_a = x(\taua)$ and $\lambda = 4 a m^2 A_0^2 / \hbar^2$, yields
\begin{equation}
\Delta u(t) = \lambda \upsilon(x) \,,
\qquad \mbox{and} \qquad
\Delta \dot{u}(t) = \frac{\hbar \lambda}{2 m A_0}  \upsilon'(x) \,,
\end{equation}
with
\begin{equation}
 \upsilon(x) = \left\{ \begin{array}{ll}
x^2 
&\mbox{for}\ x \in [0,\frac{\xi_a}{2}) \\
-x^2 + 2 \xi_a x - \frac{\xi_a^2}{2} 
&\mbox{for}\ x \in [\frac{\xi_a}{2},\xi_a) \\
\frac{\xi_a^2}{2} 
&\mbox{for}\ x \in [\xi_a,X - \xi_a) \\
-(X-x)^2 + 2 \xi_a (X-x) - \frac{\xi_a^2}{2} 
&\mbox{for}\ x \in [X - \xi_a,X - \frac{\xi_a}{2}) \\
(X-x)^2 
&\mbox{for}\ x \in [X - \frac{\xi_a}{2},X) \,.
\end{array}\right. 
\end{equation}
Defining 
and using the approximations \eref{eqn:approx-km} and \eref{eqn:approx-kmp} we find
\begin{eqnarray}
 k_1 &\approx \frac{\lambda}{a}
\left\{ \begin{array}{ll}
\Wsn^2 \frac{9\lambda^2}{10\sigma^2}\upsilon(x)^2
+ \wsn^2 \frac{9\lambda}{8R}\upsilon(x)
&\mbox{for}\ \upsilon(x) \in [0,\frac{\sigma}{\lambda}) \\
\Wsn^2 + \wsn^2 \frac{9\lambda}{8R}\upsilon(x)
&\mbox{for}\ \upsilon(x) \in [\frac{\sigma}{\lambda},\frac{2R}{\lambda}) \\
\Wsn^2 + \wsn^2 
&\mbox{for}\ \upsilon(x) \in [\frac{2R}{\lambda},\infty)  \,.
\end{array}\right. \\
 k_2 &\approx \frac{\lambda \upsilon'(x)}{4 a}
\left\{ \begin{array}{ll}
\Wsn^2 \frac{9 \lambda^2}{5 \sigma^2}\upsilon(x)
+ \wsn^2 \frac{9\lambda}{8 R}
&\mbox{for}\ \upsilon(x) \in [0,\frac{\sigma}{\lambda}) \\
-\Wsn^2 \frac{9 \sqrt{2 \pi} \sigma^3}{\lambda^3\upsilon(x)^4}  
+ \wsn^2 \frac{9\lambda}{8 R}
&\mbox{for}\ \upsilon(x) \in [\frac{\sigma}{\lambda},\frac{2R}{\lambda}) \\
-\Wsn^2  \frac{9 \sqrt{2 \pi} \sigma^3}{\lambda^3\upsilon(x)^4}  
- \wsn^2  \frac{6 R^3}{\lambda^3 \upsilon(x)^4} 
&\mbox{for}\ \upsilon(x) \in [\frac{2R}{\lambda},\infty)  \,.
\end{array}\right.
\end{eqnarray}
Labeling the different cases in ascending order by $k_{1,2}^i$ and $\upsilon^j$, where $i \in \{1,2,3\}$ and $j \in \{1,2,3,4,5\}$ we need to find the integrals
\begin{eqnarray}
K_1^{ij}(x_a,x_b) &= \int_{x_a}^{x_b} k_1^i(\upsilon^j(x)) \frac{x^3}{1+x^2} \zeta(x) \,\rmd x \\
K_2^{ij}(x_a,x_b) &= \int_{x_a}^{x_b} k_2^i(\upsilon^j(x)) x^2 \zeta(x) \,\rmd x \,.
\end{eqnarray}
The $K_1^{ij}$ can all be evaluated analytically. We also see immediately that $K_2^{i3} = 0$, since $\upsilon$ is constant for the middle interval, and find analytical expressions for all $K_2^{1j}$, $K_2^{i1}$, and $K_2^{i5}$, leaving us with $K_2^{22}$, $K_2^{24}$, $K_2^{32}$, $K_2^{34}$. For those, note that $\upsilon(X-x) = \upsilon(x)$. Hence, for $x_{a,b} \ll X$ we can approximate
\begin{equation}
\fl K_2^{i2}(x_a,x_b) + K_2^{i4}(X-x_a,X-x_b) \approx 
X^2 \zeta(X) \int_{x_a}^{x_b} k_2^i(-x^2 + 2 \xi_a x - \frac{\xi_a^2}{2}) \,\rmd x \,. \label{eqn:app-ki24-approx}
\end{equation}
The integral phase can then be obtained for the following five distinct cases:
\begin{enumerate}
\item $\lambda \xi_a^2 < 2 \sigma$ (narrow separation):
\begin{eqnarray}
\fl \Delta \phi_\mathrm{int} 
&= \sum_{n=1}^2 \Bigg[ K_n^{11}(0,\xi_a/2)
+ K_n^{12}(\xi_a/2,\xi_a)
+ K_n^{13}(\xi_a,X-\xi_a) \nnl
\fl &\bleq + K_n^{14}(X-\xi_a,X-\xi_a/2)
+ K_n^{15}(X-\xi_a/2,X) \Bigg] \nnl
\fl &\approx -\frac{3 \hbar \wsn^2 \tau^3 \Delta u_\mathrm{max}}{32 m R A_0}
\end{eqnarray}
\item $2 \sigma \leq \lambda \xi_a^2 < 4 \sigma $ (medium separation):
We then find $\lambda \upsilon(x_{1a}) = \sigma$ for
\begin{equation}
 x_{1a} = \xi_a \left(1-\sqrt{\frac{1}{2}-\frac{\sigma}{\lambda \xi_a^2}}\right) 
 > \frac{\xi_a}{2}
\end{equation}
and hence
\begin{eqnarray}
\fl \Delta \phi_\mathrm{int}
&= \sum_{n=1}^2 \Bigg[ K_n^{11}(0,\xi_a/2)
+ K_n^{12}(\xi_a/2,x_{1a})
+ K_n^{22}(x_{1a},\xi_a)
+ K_n^{23}(\xi_a,X-\xi_a) \nnl
\fl &\bleq 
+ K_n^{24}(X-\xi_a,X-x_{1a})
+ K_n^{14}(X-x_{1a},X-\xi_a/2)
\nnl \fl &\bleq 
+ K_n^{15}(X-\xi_a/2,X) \Bigg] \nnl
\fl&\approx \frac{\Wsn^2 \tau^2}{4} \Bigg[\left(6 \sqrt{\frac{\pi}{2}} \left(\frac{\sigma^3}{\Delta u_\mathrm{max}^3} - 1 \right)+\frac{1}{10}\right) \zeta(X) - \frac{2X}{3} \Bigg] \nnl
\fl&\bleq + \frac{\wsn^2 \tau^2}{4} \Bigg[\frac{9 (\Delta u_\mathrm{max}-\sigma)}{4R} \zeta(X) - \frac{3 X \Delta u_\mathrm{max}}{4 R} \Bigg]
\end{eqnarray}
\item $4 \sigma \leq \lambda \xi_a^2 < 4 R $ (medium separation):
We then find $\lambda \upsilon(x_{1b}) = \sigma$ for $x_{1b} = \sqrt{\sigma/\lambda}$ and hence
\begin{eqnarray}
\fl \Delta \phi_\mathrm{int} 
&= \sum_{n=1}^2 \Bigg[ K_n^{11}(0,x_{1b})
+ K_n^{21}(x_{1b},\xi_a/2)
+ K_n^{22}(\xi_a/2,\xi_a)
+ K_n^{23}(\xi_a,X-\xi_a) 
\nnl \fl &\bleq 
+ K_n^{24}(X-\xi_a,X-\xi_a/2)
+ K_n^{25}(X-\xi_a/2,X-x_{1b})
\nnl \fl &\bleq 
+ K_n^{15}(X-x_{1b},X) \Bigg] \nnl
\fl&\approx \frac{\Wsn^2 \tau^2}{4} \Bigg[\left(6 \sqrt{\frac{\pi}{2}} \left(1 - \frac{15 \sigma^3}{\Delta u_\mathrm{max}^3} \right)+\frac{1}{10}\right) \zeta(X) - \frac{2X}{3} \Bigg] \nnl
\fl&\bleq + \frac{\wsn^2 \tau^2}{4} \Bigg[\frac{9 \Delta u_\mathrm{max}}{8R} \zeta(X) - \frac{3 X \Delta u_\mathrm{max}}{4 R} \Bigg]
\end{eqnarray}
\item $4 R \leq \lambda \xi_a^2 < 8 R$ (large separation): This implies $4 \sigma < \lambda \xi_a^2$, hence $\lambda \upsilon(x_{1b}) = \sigma$ as before, and we find $\lambda \upsilon(x_{2a}) = 2R$ for
\begin{equation}
 x_{2a} = \xi_a \left(1-\sqrt{\frac{1}{2}-\frac{2R}{\lambda \xi_a^2}}\right) 
 > \frac{\xi_a}{2}
\end{equation}
and hence
\begin{eqnarray}
\fl \Delta \phi_\mathrm{int} 
&= \sum_{n=1}^2 \Bigg[ K_n^{11}(0,x_{1b})
+ K_n^{21}(x_{1b},\xi_a/2)
+ K_n^{22}(\xi_a/2,x_{2a})
+ K_n^{32}(x_{2a},\xi_a)
\nnl \fl &\bleq 
+ K_n^{33}(\xi_a,X-\xi_a)
+ K_n^{34}(X-\xi_a,X-x_{2a})
+ K_n^{24}(X-x_{2a},X-\xi_a/2)
\nnl \fl &\bleq 
+ K_n^{25}(X-\xi_a/2,X-x_{1b})
+ K_n^{15}(X-x_{1b},X) \Bigg] \nnl
\fl&\approx \frac{\Wsn^2 \tau^2}{4} \Bigg[\left(6 \sqrt{\frac{\pi}{2}} \left(1 - \frac{15 \sigma^3}{\Delta u_\mathrm{max}^3} \right)+\frac{1}{10}\right) \zeta(X) - \frac{2X}{3} \Bigg] \nnl
\fl&\bleq + \frac{\wsn^2 \tau^2}{4} \Bigg[\left(\frac{9 \Delta u_\mathrm{max}}{8R} - \frac{2R^3}{\Delta u_\mathrm{max}^3} - 3\right) \zeta(X) - \frac{2X}{3} \Bigg]
\end{eqnarray}
\item $8 R \leq \lambda \xi_a^2$ (large separation): Again, this implies $\lambda \upsilon(x_{1b}) = \sigma$, and we find $\lambda \upsilon(x_{2b}) = 2R$ for $x_{2b} = \sqrt{2R/\lambda}$
and hence
\begin{eqnarray}
\fl \Delta \phi_\mathrm{int} 
&= \sum_{n=1}^2 \Bigg[ K_n^{11}(0,x_{1b})
+ K_n^{21}(x_{1b},x_{2b})
+ K_n^{31}(x_{2b},\xi_a/2)
+ K_n^{32}(\xi_a/2,\xi_a)
\nnl \fl &\bleq 
+ K_n^{33}(\xi_a,X-\xi_a)
+ K_n^{34}(X-\xi_a,X-\xi_a/2)
+ K_n^{35}(X-\xi_a/2,X-x_{2b})
\nnl \fl &\bleq 
+ K_n^{25}(X-x_{2b},X-x_{1b})
+ K_n^{15}(X-x_{1b},X) \Bigg] \nnl
\fl&\approx \frac{\Wsn^2 \tau^2}{4} \Bigg[\left(6 \sqrt{\frac{\pi}{2}} \left(1 - \frac{15 \sigma^3}{\Delta u_\mathrm{max}^3} \right)+\frac{1}{10}\right) \zeta(X) - \frac{2X}{3} \Bigg] \nnl
\fl&\bleq + \frac{\wsn^2 \tau^2}{4} \Bigg[\left(\frac{30 R^3}{\Delta u_\mathrm{max}^3} + 1\right) \zeta(X) - \frac{2X}{3} \Bigg]
\end{eqnarray}
\end{enumerate}
The approximations in the last lines, respectively, are obtained under the assumption that $\xi_a \ll X$, i.\,e. $\taua \ll \tau \approx T$ and $X \approx \hbar \tau / (2 m A_0)$, making use of the approximation \eref{eqn:app-ki24-approx}, as well as $R \gg \sigma$ and $\Wsn \gg \wsn$.


\section*{References}
\begin{thebibliography}{10}

\bibitem{eppleyNecessityQuantizingGravitational1977}
Kenneth Eppley and Eric Hannah.
\newblock The necessity of quantizing the gravitational field.
\newblock {\em Found. Phys.}, 7(1-2):51--68, 1977.

\bibitem{pageIndirectEvidenceQuantum1981}
Don~N. Page and C.~D. Geilker.
\newblock Indirect {{Evidence}} for {{Quantum Gravity}}.
\newblock {\em Phys. Rev. Lett.}, 47:979--982, 1981.

\bibitem{mattinglyQuantumGravityNecessary2005}
James Mattingly.
\newblock Is {{Quantum Gravity Necessary}}?
\newblock In A.~J. Kox and Jean Eisenstaedt, editors, {\em Einstein {{Studies
  Volume}} 11. {{The Universe}} of {{General Relativity}}}, Einstein
  {{Studies}}, pages 327--338. {Birkh\"auser}, {Boston}, 2005.

\bibitem{mattinglyWhyEppleyHannah2006}
James Mattingly.
\newblock Why {{Eppley}} and {{Hannah}}'s thought experiment fails.
\newblock {\em Phys. Rev. D}, 73:064025, 2006.

\bibitem{albersMeasurementAnalysisQuantum2008}
Mark Albers, Claus Kiefer, and Marcel Reginatto.
\newblock Measurement analysis and quantum gravity.
\newblock {\em Phys. Rev. D}, 78:064051, 2008.

\bibitem{rosenfeldQuantizationFields1963}
L.~Rosenfeld.
\newblock On quantization of fields.
\newblock {\em Nucl. Phys.}, 40:353--356, 1963.

\bibitem{kibbleSemiClassicalTheoryGravity1981}
T.~W.~B. Kibble.
\newblock Is a {{Semi}}-{{Classical Theory}} of {{Gravity Viable}}?
\newblock In C.~J. Isham, R.~Penrose, and D.~W. Sciama, editors, {\em Quantum
  {{Gravity}} 2. {{A Second Oxford Symposium}}}, pages 63--80, {New York},
  1981. {Oxford University Press}.

\bibitem{penroseGravitizationQuantumMechanics2014}
Roger Penrose.
\newblock On the {{Gravitization}} of {{Quantum Mechanics}} 1: {{Quantum State
  Reduction}}.
\newblock {\em Found. Phys.}, 44(5):557--575, 2014.

\bibitem{tilloySourcingSemiclassicalGravity2016}
Antoine Tilloy and Lajos Di{\'o}si.
\newblock Sourcing semiclassical gravity from spontaneously localized quantum
  matter.
\newblock {\em Phys. Rev. D}, 93:024026, 2016.

\bibitem{carlipQuantumGravityNecessary2008}
S.~Carlip.
\newblock Is quantum gravity necessary?
\newblock {\em Class. Quantum Grav.}, 25(15):154010, 2008.

\bibitem{giuliniGravitationallyInducedInhibitions2011}
Domenico Giulini and Andr{\'e} Gro{\ss}ardt.
\newblock Gravitationally induced inhibitions of dispersion according to the
  {{Schr\"odinger}}-{{Newton}} equation.
\newblock {\em Class. Quantum Grav.}, 28(19):195026, 2011.

\bibitem{yangMacroscopicQuantumMechanics2013}
Huan Yang, Haixing Miao, Da-Shin Lee, Bassam Helou, and Yanbei Chen.
\newblock Macroscopic {{Quantum Mechanics}} in a {{Classical Spacetime}}.
\newblock {\em Phys. Rev. Lett.}, 110(17):170401, 2013.

\bibitem{grossardtOptomechanicalTestSchrodingerNewton2016}
Andr{\'e} Gro{\ss}ardt, James Bateman, Hendrik Ulbricht, and Angelo Bassi.
\newblock Optomechanical test of the {{Schr\"odinger}}-{{Newton}} equation.
\newblock {\em Phys. Rev. D}, 93:096003, 2016.

\bibitem{giuliniSchrodingerNewtonEquationNonrelativistic2012}
Domenico Giulini and Andr{\'e} Gro{\ss}ardt.
\newblock The {{Schr\"odinger}}-{{Newton}} equation as a nonrelativistic limit
  of self-gravitating {{Klein}}-{{Gordon}} and {{Dirac}} fields.
\newblock {\em Class. Quantum Grav.}, 29(21):215010, 2012.

\bibitem{bahramiSchrodingerNewtonEquationIts2014}
Mohammad Bahrami, Andr{\'e} Gro{\ss}ardt, Sandro Donadi, and Angelo Bassi.
\newblock The {{Schr\"odinger}}-{{Newton}} equation and its foundations.
\newblock {\em New J. Phys.}, 16:115007, 2014.

\bibitem{boseSpinEntanglementWitness2017}
Sougato Bose, Anupam Mazumdar, Gavin~W. Morley, Hendrik Ulbricht, Marko Toro{\v
  s}, Mauro Paternostro, Andrew~A. Geraci, Peter~F. Barker, M.~S. Kim, and
  Gerard Milburn.
\newblock Spin {{Entanglement Witness}} for {{Quantum Gravity}}.
\newblock {\em Phys. Rev. Lett.}, 119:240401, 2017.

\bibitem{marlettoGravitationallyInducedEntanglement2017}
C.~Marletto and V.~Vedral.
\newblock Gravitationally {{Induced Entanglement}} between {{Two Massive
  Particles}} is {{Sufficient Evidence}} of {{Quantum Effects}} in {{Gravity}}.
\newblock {\em Phys. Rev. Lett.}, 119(24):240402, December 2017.

\bibitem{kafriClassicalChannelModel2014}
D.~Kafri, J.~M. Taylor, and G.~J. Milburn.
\newblock A classical channel model for gravitational decoherence.
\newblock {\em New J. Phys.}, 16:065020, 2014.

\bibitem{grossardtAccelerationNoiseConstraints2020}
Andr{\'e} Gro{\ss}ardt.
\newblock Acceleration noise constraints on gravity-induced entanglement.
\newblock {\em Phys. Rev. A}, 102(4):040202(R), 2020.

\bibitem{giuliniCentreofmassMotionMultiparticle2014}
Domenico Giulini and Andr{\'e} Gro{\ss}ardt.
\newblock Centre-of-mass motion in multi-particle {{Schr\"odinger}}-{{Newton}}
  dynamics.
\newblock {\em New J. Phys.}, 16:075005, 2014.

\bibitem{grossardtEffectsNewtonianGravitational2016}
Andr{\'e} Gro{\ss}ardt, James Bateman, Hendrik Ulbricht, and Angelo Bassi.
\newblock Effects of {{Newtonian}} gravitational self-interaction in
  harmonically trapped quantum systems.
\newblock {\em Sci. Rep.}, 6:30840, 2016.

\bibitem{colellaObservationGravitationallyInduced1975}
R.~Colella, A.~W. Overhauser, and S.~A. Werner.
\newblock Observation of {{Gravitationally Induced Quantum Interference}}.
\newblock {\em Phys. Rev. Lett.}, 34(23):1472--1474, 1975.

\bibitem{colinCrucialTestsMacrorealist2016}
Samuel Colin, Thomas Durt, and Ralph Willox.
\newblock Crucial tests of macrorealist and semiclassical gravity models with
  freely falling mesoscopic nanospheres.
\newblock {\em Phys. Rev. A}, 93:062102, 2016.

\bibitem{grossardtApproximationsFreeEvolution2016}
Andr{\'e} Gro{\ss}ardt.
\newblock Approximations for the free evolution of self-gravitating quantum
  particles.
\newblock {\em Phys. Rev. A}, 94:022101, 2016.

\bibitem{oliveReviewParticlePhysics2014}
K.A. Olive.
\newblock Review of {{Particle Physics}}.
\newblock {\em Chinese Phys. C}, 38(9):090001, 2014.

\bibitem{gaoParameterizationTemperatureDependence1999}
H.~X. Gao and L.-M. Peng.
\newblock Parameterization of the temperature dependence of the
  {{Debye}}-{{Waller}} factors.
\newblock {\em Acta Crystallogr. Sect. A}, 55:926--932, 1999.

\bibitem{searsDebyeWallerFactorElemental1991}
V.~F. Sears and S.~A. Shelley.
\newblock Debye-{{Waller Factor}} for {{Elemental Crystals}}.
\newblock {\em Acta Crystallogr. Sect. A}, 47:441--446, 1991.

\bibitem{iweCoulombPotentialsSpherical1982}
H.~Iwe.
\newblock Coulomb {{Potentials Between Spherical Heavy Ions}}.
\newblock {\em Z. Phys.}, 304(4):347--361, 1982.

\bibitem{hatifiRevealingSelfgravitySternGerlach2020}
Mohamed Hatifi and Thomas Durt.
\newblock Revealing self-gravity in a {{Stern}}-{{Gerlach Humpty}}-{{Dumpty}}
  experiment.
\newblock {\em arXiv:2006.07420 [gr-qc, physics:quant-ph]}, June 2020.

\bibitem{seligDropTowerMicrogravity2010}
Hanns Selig, Hansj{\"o}rg Dittus, and Claus L{\"a}mmerzahl.
\newblock Drop {{Tower Microgravity Improvement Towards}} the {{Nano}}-g
  {{Level}} for the {{MICROSCOPE Payload Tests}}.
\newblock {\em Microgravity Sci. Technol.}, 22(4):539--549, 2010.

\bibitem{altamiranoGravityNotPairwise2018}
Natacha Altamirano, Paulina {Corona-Ugalde}, Robert~B. Mann, and Magdalena
  Zych.
\newblock Gravity is not a pairwise local classical channel.
\newblock {\em Class. Quantum Grav.}, 35(14):145005, 2018.

\bibitem{donadiUndergroundTestGravityrelated2021}
Sandro Donadi, Kristian Piscicchia, Catalina Curceanu, Lajos Di{\'o}si,
  Matthias Laubenstein, and Angelo Bassi.
\newblock Underground test of gravity-related wave function collapse.
\newblock {\em Nat. Phys.}, 17(1):74--78, 2021.

\end{thebibliography}
\end{document}